\documentclass[11pt,a4paper]{article}
\usepackage[noadjust]{cite}

\usepackage[margin=2.8cm,bottom=3.5cm]{geometry}

\usepackage{amsmath, amssymb}
\usepackage{color}
\pdfoutput=1 

\usepackage{graphicx} 
\usepackage{subcaption}
\usepackage[T1]{fontenc}

\title{Kink-antikink collision in the supersymmetric $\phi^4$ model}

\author{
        Jo\~ao G. F. Campos \\
        Departamento de F\'isica, Universidade Federal da Para\'iba,\\
        Jo\~ao Pessoa - PB - 58051-970, Brazil\\
        joaogfc@gmail.com
            \and
        Azadeh Mohammadi\\
        Departamento de F\'isica, Universidade Federal de Pernambuco,\\
        Av. Prof. Moraes Rego, 1235, Recife - PE - 50670-901, Brazil\\
        azadeh.mohammadi@ufpe.br
}

\begin{document} 
\maketitle

\begin{abstract}

This paper investigates a model containing $\phi^4$ kinks interacting with fermions. The fermion back-reaction is included in the equations of motion, which affects the kink-antikink collisions. We show that the fermion field generates a force that can be either attractive or repulsive. Moreover, we investigate three different scenarios, which exhibit a wide variety of behaviors, including the usual scenarios observed in the $\phi^4$ model as well as the formation of two oscillons, reflection without contact, one-bounce resonance windows, and the creation of kink-antikink pairs. We also find evidence that the fermion field can store part of the energy responsible for the energy exchange mechanism.

\end{abstract}

\section{Introduction}

Linear and nonlinear wave field equations are ubiquitous in physics, with applications in almost all research fields. One fascinating feature of the nonlinear ones is that they may exhibit solitary waves and solitons. The solitary waves are localized structures that propagate without losing their shape. Solitons have the additional property that, after a collision, the initial constituents reappear unchanged, except for a phase shift. Examples of localized structures are kinks, vortices, skyrmions, and monopoles \cite{rajaraman1982solitons, manton2004topological}.

Kinks and antikinks, their $x\to -x$ counterparts, are localized solutions that appear in systems that effectively have one spatial dimension. In higher dimensions, the kinks can be extended, forming domain walls. They have applications in many physical systems such as polymers \cite{su1979solitons}, Josephson junctions \cite{ustinov1998solitons}, DNA \cite{yakushevich2006nonlinear}, deformations in graphene \cite{yamaletdinov2017kinks} and surface displacement in shallow liquids \cite{denardo1990observation}. 
The two prototypical models which describe kinks are the sine-Gordon and $\phi^4$. The former is integrable and exhibits solitons, while the latter is non-integrable and exhibits solitary waves. Thus, the behavior of $\phi^4$ kinks is much richer because their interactions are highly inelastic. Since the triplet of seminal works on kink-antikink collisions \cite{campbell1983resonance, peyrard1983kink, campbell1986kink}, it is known that, for non-integrable models, a kink-antikink collision can exhibit resonance windows. These are intervals of the initial kink's relative velocity where they separate after multiple bounces. The explanation given for this phenomenon is called the resonant energy exchange mechanism. It states that, after the first collision, the translational energy is converted into vibrational energy. The vibrational energy does not dissipate and can be recovered in subsequent collisions, allowing the system to finally separate. The set of initial velocities where resonance windows occur forms a fractal structure. It means that the system is chaotic and consequently has a high sensitivity to initial conditions. Such complex behavior can only happen for non-integrable models.

A few examples of recent investigations involving kink-antikink interactions consider wobbling kinks \cite{izquierdo2021scattering, campos2021wobbling}, kinks with quasinormal modes \cite{dorey2018resonant, campos2020quasinormal}, long-range \cite{christov2019long, christov2019kink, christov2021kink, campos2021interaction} and short-range \cite{bazeia2021semi} kinks. Other interesting works  involve hyperbolic models \cite{bazeia2018scattering, bazeia2019kink, bazeia2020oscillons}, models with Bogomol'nyi-Prasad-Sommerfield (BPS) impurities \cite{adam2019spectral, adam2019solvable, adam2019phi, adam2020kink}, interaction with boundaries \cite{arthur2016breaking, dorey2017boundary, lima2019boundary} and an interpolation between the $\phi^4$ and sine-Gordon models \cite{dorey2021resonance}. Moreover, advances in the application of collective coordinates to kink-antikink collisions were able to give a more quantitative description of the resonant energy exchange mechanism \cite{manton2021collective, manton2021kink,adam2022relativistic}. 

Solitary waves interactions for multiple scalar fields have been investigated in the literature \cite{halavanau2012resonance, alonso2018reflection, alonso2020non, alonso2021kink}. However, studies involving collisions of solitary waves composed of both scalar and fermionic fields are still lacking. 
The interaction of topological structures with fermions leads to many novel properties such as fractional quantum numbers \cite{jackiw1981zero, goldstone1981fractional}, and domain wall and string superconductivity \cite{vachaspati2006kinks,vilenkin2000cosmic}. The system composed of a fermion interacting with a single kink via a Yukawa coupling can be solved analytically if the back-reaction is ignored \cite{charmchi2014complete, charmchi2014massive}. As discussed below, these results are important because the back-reaction vanishes at the limit of a very massive kink. However, in general, the effect of the back-reaction can be significant. It can be self-consistently included, as shown in Refs. \cite{amado2017coupled, klimashonok2019fermions} for kinks, and in Ref. \cite{perapechka2018soliton} for baby-skyrmions. Interestingly, two kinks or two baby-skyrmions can bind due to the presence of a fermion field \cite{perapechka2020kinks, perapechka2019fermion}.

Here, we will consider collisions between a kink and an antikink interacting with a fermion field, including the fermion back-reaction on the kinks. In \cite{chu2008fermions,brihaye2008remarks} fermions interacting with a kink-antikink configuration have been investigated, although with a frozen scalar field configuration, i.e., without evolving in time and without fermion back-reaction. In \cite{gibbons2007fermions, saffin2007particle}, the authors studied the kink-antikink collision, but without the back-reaction as well. More recently, we have studied fermions interacting with another dynamic scalar field with the same assumption \cite{campos2020fermion, campos2021fermions}. Therefore, it is essential to study the effect created by the fermion on the scalar field evolution, which is lacking in the literature.

In section \ref{sec_model}, we introduce briefly a supersymmetric extension of the $\phi^4$ model, which is the simplest way to add a fermion field with minimum free parameters. We will define symmetric configurations of interest and compute the inter-fermion force. In section \ref{sec_stab}, we will briefly discuss the stability equation of the coupled system. In section \ref{sec_results}, we will discuss the numerical results for kink-antikink collision and the corresponding fermion field configurations. Finally, we will summarize our findings in the last section.

\section{Model}
\label{sec_model}

We are interested in the field theory in (1+1) dimensions described by the following action
\begin{equation}
S=\int d^2y\left(\frac{1}{2}\partial^\mu\phi\partial_\mu\phi+\frac{1}{2}i\bar{\Psi}\gamma^\mu\partial_\mu\Psi+\frac{1}{2}F^2+W_\phi F-\frac{1}{2}W_{\phi\phi}\bar{\Psi}\Psi\right),
\end{equation}
which can be obtained from the minimal supersymmetric model studied in Refs. \cite{shifman1999anomaly, shizuya2004superfield}. The model is invariant under a supersymmetry transformation which guarantees the existence of a supercurrent and a supercharge. For more detail on supersymmetry algebra and supersymmetric solitons, we refer the reader to the two references cited above.

Substituting the equation of motion for the auxiliary field $F=-W_\phi$, and choosing
\begin{equation}
W_\phi=\sqrt{\frac{\lambda}{2}}\left(\phi^2-\frac{m^2}{\lambda}\right),
\end{equation} 
it becomes
\begin{equation}
S=\int d^2y~\left(\frac{1}{2}\partial^\mu\phi\partial_\mu\phi+\frac{1}{2}i\bar{\Psi}\gamma^\mu\partial_\mu\Psi-\sqrt{\frac{\lambda}{2}}\phi\bar{\Psi}\Psi-V(\phi)\right),
\end{equation}
with $V(\phi)=\frac{1}{2} W_\phi^2$. The above action consists of the $\phi^4$ model coupled with a Dirac field via a Yukawa coupling, ensuring the supersymmetric form. The equations of motion obtained from the action should be accompanied by the normalization condition
\begin{equation}
\int_{-\infty}^\infty\Psi^\dag(y,\tau)\Psi(y,\tau)dy=1.
\end{equation}

Now, we perform the usual rescaling to simplify the $\phi^4$ potential, besides working with dimensionless parameters. We define the kink's value at infinity $\phi_0=m\lambda^{-\frac{1}{2}}$, and set $\phi\to\phi_0\chi$ and $y^\mu\to \phi_0^{-1}\lambda^{-\frac{1}{2}}x^\mu$. To maintain the normalization condition we also set $\Psi\to\phi_0^{\frac{1}{2}}\lambda^{\frac{1}{4}}\psi$. This way we work with the following dimensionless lagrangian density 
\begin{equation}
{\cal L}=\frac{\phi_0^2}{2}\partial^\mu\chi\partial_\mu\chi+\frac{1}{2}i\bar{\psi}\gamma^\mu\partial_\mu\psi-\frac{1}{\sqrt{2}}\chi\bar{\psi}\psi-\frac{\phi_0^2}{4}\left(\chi^2-1\right)^2.
\end{equation}
supplemented by the normalization condition
\begin{equation}
\int_{-\infty}^\infty\psi^\dag(x,t)\psi(x,t)dx=1.
\end{equation}
The equations of motion for this model are
\begin{align}
\label{eompsi}
&i\gamma^\mu\partial_\mu\psi-\sqrt{2}\chi\psi=0,\\
&\partial_\mu\partial^\mu\chi+\chi\left(\chi^2-1\right)+\frac{1}{\sqrt{2}\phi_0^2}\bar{\psi}\psi=0.
\label{eomchi}
\end{align}
The first equation is the Dirac equation for a fermion field coupled to the scalar field $\chi$. The second is the equation of motion for the rescaled $\phi^4$ model plus the term with the fermion field giving rise to the fermion back-reaction. Looking at the above equations of motion, one could be tempted to rescale the fermion field to eliminate the parameter $\phi_0$. The problem is that this rescaling would modify the normalization condition. In fact, it is not difficult to see that $\phi_0$ quantifies the strength of the back-reaction.  
Notice that the Dirac equation has fixed parameters because it does not depend on $\phi_0$. In the limit when the parameter $\phi_0$ is very large, the back-reaction term in eq.~(\ref{eomchi})  vanishes. Therefore, the approximation where the fermion back-reaction is neglected is justified for a very large $\phi_0$, weak-coupling regime. In other words, a very massive kink does not feel the fermion. However, for a small $\phi_0$, large-coupling regime, the impact of the fermion back-reaction on the scalar field can be considerably large. This is the regime we are most interested in.

\subsection{Bound states}

To move forward, we need to choose a representation for Dirac matrices. We choose $\gamma^0=\sigma_1$, and $\gamma^1=i\sigma_3$, where $\sigma_i$ are the Pauli matrices. Writing $\psi=e^{-iEt}(u\;v)^T$, with real $u$ and $v$, besides considering the scalar field to be static, leads to the coupled equations of motion as follows
\begin{align}
\label{eq_stat1}
&Eu+v^\prime-\sqrt{2}\chi v=0,\\
\label{eq_stat2}
&Ev-u^\prime-\sqrt{2}\chi u=0,\\
\label{eq_stat3}
&\chi^{\prime\prime}-\chi(\chi^2-1)-\frac{\sqrt{2}}{\phi_0^2}uv=0.
\end{align}
These equations are supplemented by the normalization condition
\begin{equation}
\int dx\left(u^2+v^2\right)=1.
\end{equation} 
It is instructive to derive the fermion zero mode solution, which is required by the index theorem. Setting $E=0$, eqs. \ref{eq_stat1}-\ref{eq_stat2} can be solved in the form
\begin{align}
u&=a_1e^{-\sqrt{2}\int^x \chi(x^\prime)dx^\prime}\\
v&=a_2e^{\sqrt{2}\int^x \chi(x^\prime)dx^\prime}
\end{align}
For a kink solution $\chi(x\to\pm\infty)=\pm 1$. Thus, $v$ diverges, unless $a_2=0$. With the same argument, for an antikink, $a_1=0$. In both cases, the product $uv=0$, and the back-reaction vanishes, meaning that the fermion does not deform the $\phi^4$ kink solution
\begin{equation}
\chi_k(x)=\tanh\left(\frac{x}{\sqrt{2}}\right).
\end{equation}
This allows one to find the normalized zero mode solution
\begin{equation}
\psi_0(x)=\sqrt{\frac{3}{4\sqrt{2}}}\begin{pmatrix}
\text{sech}^2\left(\frac{x}{\sqrt{2}}\right)\\
0
\end{pmatrix}.
\end{equation}
For the antikink $\chi_{\bar{k}}(x)=-\chi_k(x)$, we have instead
\begin{equation}
\psi_{\bar{0}}(x)=\sqrt{\frac{3}{4\sqrt{2}}}\begin{pmatrix}
0\\
\text{sech}^2\left(\frac{x}{\sqrt{2}}\right)
\end{pmatrix}.
\end{equation}
As both kink and antikink solutions, $\chi_k$ and $\chi_{\bar{k}}$, obey the BPS equation, they preserve half of the supersymmetry \cite{shifman1999anomaly}.

To find other solutions, eqs. (\ref{eq_stat1})-(\ref{eq_stat3}) can be solved self-consistently using relaxation methods, as done in Refs. \cite{amado2017coupled, klimashonok2019fermions}. Here, we solve the system in the interval $0.0<x<50.0$ uniformly separated by 2500 points. The center of the kink is at $x=0$, and the solution for negative $x$ is obtained using the solution's parity properties. Then, the derivatives are discretized using a five-point stencil approximation. The resulting nonlinear system is solved by utilizing the optimize.root function in Python's SciPy library. 

\begin{figure}[tbp]
\centering
   \includegraphics[width=0.65\linewidth]{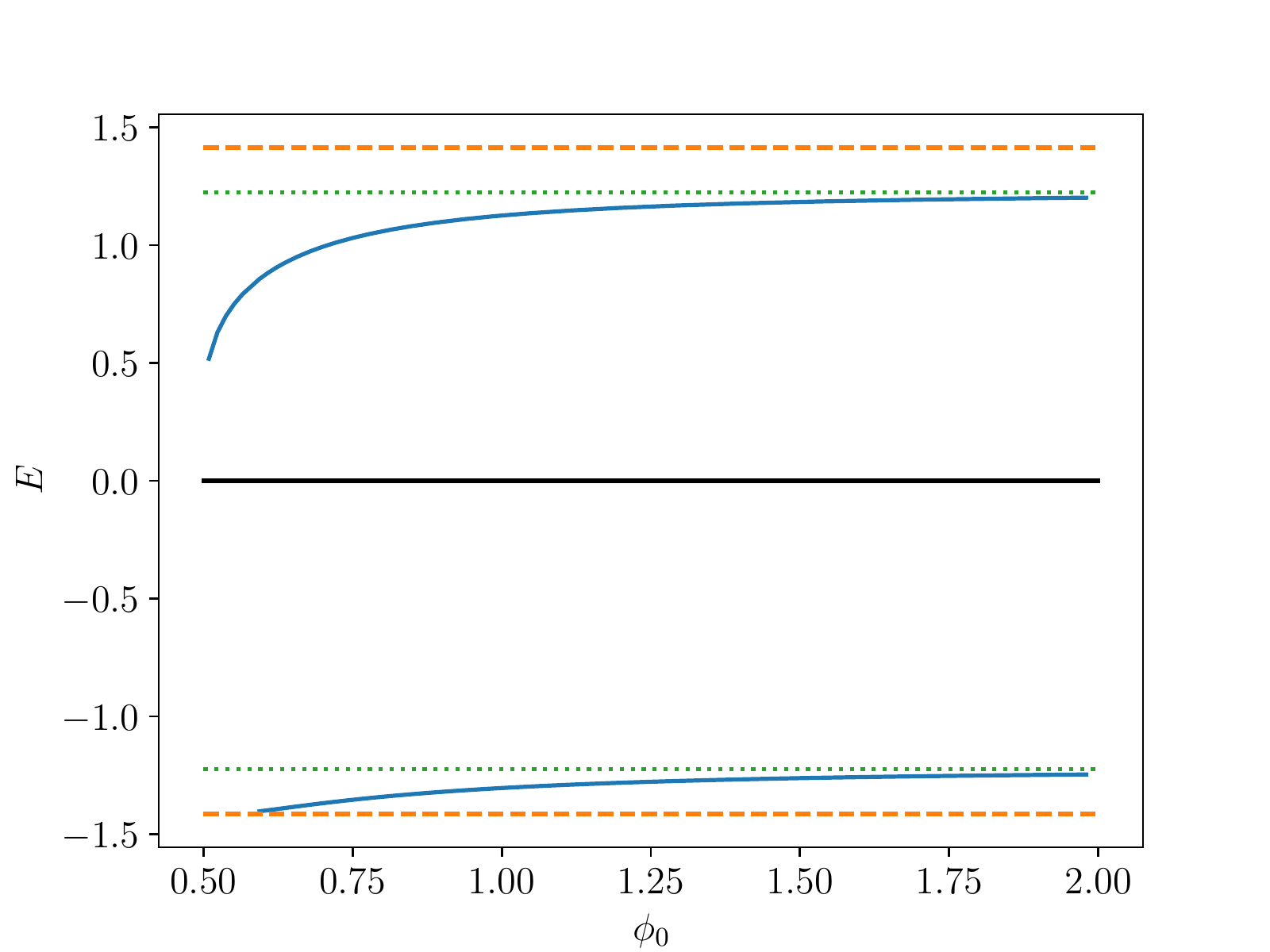}
   \caption{Energy eigenvalues (solid curves) as a function of the parameter $\phi_0$. Dashed lines correspond to the threshold of the continuum spectrum $E=\sqrt{2}$. Dotted lines correspond to the analytical value obtained in the absence of back-reaction $E=\sqrt{3/2}$. This figure is a reproduction of results obtained in \cite{amado2017coupled}.}
   \label{fig_en}
\end{figure}

In the supersymmetric case, besides the zero mode, there are also two other bound states with positive and negative energy. They are related by the energy-reflection symmetry when there is no back-reaction. However, the back-reaction breaks this symmetry. Unlike the fermion zero mode, these solutions deform the kink and depend on $\phi_0$. We denote them by $\chi_{k,\pm 1}(x;\phi_0)$ and $\psi_{\pm 1}(x,t;\phi_0)$. There is a phase freedom in $\psi_{\pm 1}(x,t;\phi_0)$, which we fix by choosing the real solution whose second component is positive. We also take $\chi_{\bar{k},\pm1}(x;\phi_0)=-\chi_{k,\pm1}(x;\phi_0)$ and $\psi_{\pm \bar{1}}(x,t;\phi_0)=i\sigma_2\psi_{\pm 1}(x,t;\phi_0)$, where the bar above the subscript corresponds to an antikink. It is easy to show that they are also solutions to the equations of motion. The discrete energy eigenvalues as a function of $\phi_0$ are shown in Fig.~\ref{fig_en}. Both positive and negative energy eigenvalues are smaller than $\pm\sqrt{3/2}$, respectively. These constants are the eigenvalues obtained without back-reaction, which are approached as $\phi_0$ increases. We do not plot the positive energy eigenvalues for $\phi_0\lesssim 0.51$ because the numerical method that we employed to compute them does not reach the desired accuracy\footnote{In Ref.~\cite{amado2017coupled}, the authors had similar accuracy issues near the same range of $\phi_0$ values.}. On the other hand, the negative energy state enters the continuum at $\phi_0\simeq0.59$

To study the system's dynamics, we also need boosted solutions, which can be obtained in the usual way. For the fermion field $\psi=(\psi_1\;\psi_2)^T$, a boost with velocity $v_i$ reads
\begin{align}
\psi_1^\prime(x,t)&=\cosh(\nu/2)\psi_1(x^\prime,t^\prime)-i\sinh(\nu/2)\psi_2(x^\prime,t^\prime)\\
\psi_2^\prime(x,t)&=i\sinh(\nu/2)\psi_1(x^\prime,t^\prime)+\cosh(\nu/2)\psi_2(x^\prime,t^\prime),
\end{align}
where we defined the rapidity $\nu=\tanh^{-1}(v_i)$, and the transformed coordinates $x^\prime=x-v_it$ and $t^\prime=t-v_ix$. A boosted solution will be identified by an explicit dependence on $t$ and $v_i$. For example, $\chi_k(x,t;v_i)$ denotes a boosted kink.

\subsection{Symmetric collisions}
\label{sec_sym}
First, let us focus on symmetric collisions where before the kinks superpose, the fields can be approximately described by
\begin{align}
\label{eq_z1}
\chi(x,t)&=\chi_k(x+x_0;v_i)-\chi_k(x-x_0;-v_i)-1,\\
\psi(x,t)&=\frac{1}{\sqrt{2}}\psi_0(x+x_0;v_i)+\frac{A}{\sqrt{2}}\psi_{\bar{0}}(x-x_0;-v_i),
\label{eq_z2}
\end{align}
where $A$ is a complex phase in general and $\psi_{0}(\psi_{\bar{0}})$ is the fermion zero mode bound to a kink(antikink). We plan to find the value of $A$ for which the fermion will have the largest impact on the scalar field for a fixed value of $\phi_0$. To simplify our analysis we fix $v_i=0$. Then
\begin{equation}
\psi(x,t)=\sqrt{\frac{3}{8\sqrt{2}}}\begin{pmatrix}
\text{sech}^2\left(\frac{x+x_0}{\sqrt{2}}\right)\\
A\,\text{sech}^2\left(\frac{x-x_0}{\sqrt{2}}\right)
\end{pmatrix}.
\end{equation}
The back-reaction is proportional to
\begin{equation}
\bar{\psi}(x,t)\psi(x,t)=\frac{3}{8\sqrt{2}}\text{sech}^2\left(\frac{x+x_0}{\sqrt{2}}\right)\text{sech}^2\left(\frac{x-x_0}{\sqrt{2}}\right)(A+A^*).
\end{equation}
The absolute value of this expression is largest when $A=\pm1$. We will study collisions for these two values of $A$.

The second example is the symmetric configuration where the fermion is in the excited state with positive energy. It reads
\begin{align}
\label{eq_e1}
\chi(x,t)&=\chi_{k,+1}(x+x_0;\sqrt{2}\phi_0,v_i)-\chi_{k,+1}(x-x_0;\sqrt{2}\phi_0,-v_i)-1,\\
\psi(x,t)&=\frac{1}{\sqrt{2}}\psi_{+1}(x+x_0,t;\sqrt{2}\phi_0,v_i)+\frac{A}{\sqrt{2}}\psi_{+\bar{1}}(x-x_0,t;\sqrt{2}\phi_0,-v_i).
\label{eq_e2}
\end{align}
It is necessary to multiply the parameter $\phi_0$ by $\sqrt{2}$ to compensate for the reduction in the strength of the back-reaction. Fixing $v_i=0$ and writing $\psi_{+1}(x,t;\sqrt{2}\phi_0)=e^{-iEt}(u(x)\;v(x))^T$, the back-reaction is proportional to
\begin{equation}
\bar{\psi}(x,t)\psi(x,t)=\mathcal{B}_{1}+\mathcal{B}_{2},
\end{equation}
with
\begin{align}
&\mathcal{B}_{1}=(u(x+x_0)v(x+x_0)-u(x-x_0)v(x-x_0)),\\
&\mathcal{B}_{2}=\frac{(A+A^*)}{2}(v(x-x_0)v(x+x_0)-u(x-x_0)u(x+x_0)).
\end{align}
The first term quantifies the fermion back-reaction on the kink at the same position. On the other hand, the second one describes the back-reaction due to the overlap, which depends on the $A$ choice. Again, the fermion back-reaction is maximized for $A=\pm1$. In what follows, for brevity, we will analyze only the positive sign for the excited states. None of the kink-antikink configurations discussed above is BPS and, therefore, breaks supersymmetry.

\subsection{Energy-momentum tensor and force}

Due to the translational invariance of the model, there are four conserved currents described by the energy-momentum tensor. One can show that the tensor is given by
\begin{equation}
T^{\mu\nu}=T^{\mu\nu}_{\text{scalar}}+T^{\mu\nu}_{\text{Dirac}},
\end{equation}
where
\begin{equation}
\frac{T^{\mu\nu}_{\text{scalar}}}{\phi_0^2}=\partial^\mu\chi\partial^\nu \chi-\eta^{\mu\nu}\left(\frac{1}{2}\partial^\sigma\chi\partial_\sigma\chi-\frac{1}{4}(\chi^2-1)^2\right),
\end{equation}
and
\begin{equation}
T^{\mu\nu}_{\text{Dirac}}=\frac{i}{4}\left[\bar{\psi}\gamma^\mu\left(\partial^\nu\psi\right)-\left(\partial^\nu\bar{\psi}\right)\gamma^\mu\psi\right].
\end{equation}

Suppose that we have a kink-antikink configuration where the two are symmetrically placed around the origin and bound to fermions. Examples are given in eqs.~(\ref{eq_z1}) and (\ref{eq_z2}) as well as in eqs.~(\ref{eq_e1}) and (\ref{eq_e2}). We can easily compute the force between the two composite systems for such cases. The momentum on the left side is
\begin{equation}
P=\int_{-\infty}^0 dxT^{01}(x),
\end{equation}
and the force is
\begin{equation}
F=\frac{dP}{dt}=\int_{-\infty}^0 dx \partial_tT^{01}(x)=-\int_{-\infty}^0 dx \partial_xT^{11}(x)=-T^{11}(0).
\end{equation}
If we consider eqs.~(\ref{eq_z1}) and (\ref{eq_z2}) with $v_i=0$ and $A=\pm1$, the scalar field contribution is the usual one $F_{\text{scalar}}\simeq16\phi_0^2e^{-2\sqrt{2}x_0}$, while the fermion field causes a force $F_{\text{Dirac}}\simeq\pm 6 e^{-2\sqrt{2}x_0}$, respectively. Therefore, $A=+1$ corresponds to the attractive case and $A=-1$ to the repulsive one. The total force is $F=F_{\text{scalar}}+F_{\text{Dirac}}$. One can see that, the fermion force surpasses the scalar field one as the parameter $\phi_0$ decreases. 

We can also compute the force for the configuration described by eqs.~(\ref{eq_e1}) and (\ref{eq_e2}) with $v_i=0$ and $A=+1$. We write the excited state as $\psi_{+1}(x,t;\phi_0)=e^{-iEt}(u(x)\;v(x))^T$. It is easy to show that the asymptotic form of the state for $x\gg1$ is
\begin{equation}
\psi_{+1}(x,t;\phi_0)\simeq e^{-iEt}
\begin{pmatrix}
\frac 1E(\sqrt{2}+\kappa)v(x)\\
v(x)
\end{pmatrix},
\end{equation}
with $v(x)= Ce^{-\kappa x}$ knowing that $\kappa\equiv\sqrt{2-E^2}$ and the constant $C$ is real. Besides that, the parity symmetry implies $v(x)= v(-x)$ and $u(x)= -u(-x)$. Therefore, the force can be approximated by
\begin{equation}
F=\frac12[u(x_0)+v(x_0)][-u^\prime(x_0)+v^\prime(x_0)]\simeq C^2\, \frac{\kappa^2(\sqrt{2}+\kappa)}{E^2}\, e^{-2\kappa  x_0},
\end{equation}
Therefore, we find that the force is again attractive.

\section{Stability equation}
\label{sec_stab}

It is also important to look at the stability of the fermion-kink system. To do so, we write $\chi(x,t)=\chi_k(x)+\eta(x,t)$, $\dot{\chi}(x,t)=\dot{\eta}(x,t)\equiv\zeta(x,t)$ and
\begin{equation}
\label{eq_pert}
\psi(x,t)=e^{-iEt}\left[\begin{pmatrix}
u(x)\\
v(x)
\end{pmatrix}+\begin{pmatrix}
\alpha_1(x,t)+i\alpha_2(x,t)\\
\beta_1(x,t)+i\beta_2(x,t)\end{pmatrix}\right],
\end{equation}
where $\eta$, $\zeta$, $\alpha_1$, $\alpha_2$, $\beta_1$ and $\beta_2$ are small and real. The second column matrix in eq.~(\ref{eq_pert}) will be denoted by $\delta\psi$. Substituting the above expressions in the equations of motion leads to the following linear system
\begin{equation}
\begin{pmatrix}
\dot{\eta}\\
\dot{\zeta}\\
\dot{\alpha}_1\\
\dot{\alpha}_2\\
\dot{\beta}_1\\
\dot{\beta}_2
\end{pmatrix}=
\begin{pmatrix}
0&1&0&0&0&0\\
\frac{d^2}{dx^2}-V^{\prime\prime}(\chi_k)&0&-\frac{\sqrt{2}}{\phi_0^2}v&0&-\frac{\sqrt{2}}{\phi_0^2}u&0\\
0&0&0&-E&0&-\frac{d}{dx}+\sqrt{2}\chi_k\\
-\sqrt{2}v&0&E&0&\frac{d}{dx}-\sqrt{2}\chi_k&0\\
0&0&0&\frac{d}{dx}+\sqrt{2}\chi_k&0&-E\\
-\sqrt{2}u&0&-\frac{d}{dx}-\sqrt{2}\chi_k&0&E&0
\end{pmatrix}
\begin{pmatrix}
\eta\\
\zeta\\
\alpha_1\\
\alpha_2\\
\beta_1\\
\beta_2
\end{pmatrix}
\end{equation}
The solution is stable if there are no eigenvalues $\lambda$ with a positive real part. Moreover, an oscillatory solution is purely imaginary and comes with its complex conjugate. Therefore, the number of modes is half the number of eigenvalues. To find the spectrum, we discretize the derivatives using a five-point stencil approximation, and the resulting matrix is diagonalized using the NumPy library in Python. This method has a few difficulties that make the analysis very challenging. First, the operator is not Hermitian. This means that the resulting matrix is not symmetric, and its eigenvectors are not orthogonal. Thus, they cannot be interpreted as independent internal modes. The other issues will be explained below.

For the fermions at the zero mode, $E=0$ and
\begin{equation}
\begin{pmatrix}
u\\
v
\end{pmatrix}=\psi_0,
\end{equation}
one can find two modes with $\lambda=0$ analytically. The first one is given by the derivative of the unperturbed solution with respect to $x$
\begin{align}
&\eta_{0,s}=\mathcal{N}\text{sech}^2(x/\sqrt{2}),\quad\zeta_{0,s}=0,\\
&\delta\psi_{0,s}=\mathcal{N}\sqrt{\frac{3}{4\sqrt{2}}}\begin{pmatrix}-2\,\text{sech}^2(x/\sqrt{2})\tanh(x/\sqrt{2})\\
0\end{pmatrix}.
\end{align}
This solution is guaranteed to solve the linearized system by the translational symmetry of the model. From this construction, it should be clear that this mode receives no back-reaction, as well. The second mode with $\lambda=0$ is easily found in the form
\begin{equation}
\eta_{0,f}=0,\quad\zeta_{0,f}=0,\quad\delta\psi_{0,f}=\psi_0.
\end{equation}
Then, the other modes can be computed numerically. We found three vibrational modes with purely imaginary eigenvalues and no $\phi_0$ dependence. Thus, all three modes have vanishing back-reaction
with absolute value $|\lambda|=\sqrt{3/2}$ within our numerical precision. 
We did not find any vibrational mode with non-vanishing back-reaction using this numerical method. The spectrum of the linear matrix is closely related to the combined spectrum of scalar perturbations around an isolated kink and fermionic bound states on a kink background without back-reaction. In this case, both excitation frequencies are equal as the model is supersymmetric. 

In general, our numerical analysis shows that the composite system is stable because we did not find any eigenvalues with a positive real part, even for a fermion in the excited state. However, we avoid a more detailed stability analysis of the latter case. The main issue is that the excited state is close to the threshold between discrete and continuum eigenvalues, making it difficult to distinguish between the discrete vibrational modes and continuum states.


\section{Results}
\label{sec_results}

\subsection{Dynamics}

To evolve the full equations of motion in time, we need to define the real and imaginary components of the fermionic field as follows $\psi=(u_1+iu_2\;v_1+iv_2)^T$. Thus, the equations of motion, eqs.~(\ref{eompsi}) and (\ref{eomchi}), result in
\begin{align}
\ddot{\chi}&=\partial_x^2\chi-\chi(\chi^2-1)-\frac{\sqrt{2}}{\phi_0^2}(u_1v_1+u_2v_2),\\
\dot{u}_1&=-\partial_xv_2+\sqrt{2}\chi v_2,\\
\dot{u}_2&=\partial_xv_1-\sqrt{2}\chi v_1,\\
\dot{v}_1&=\partial_xu_2+\sqrt{2}\chi u_2,\\
\dot{v}_2&=-\partial_xu_1-\sqrt{2}\chi u_1.
\end{align}
Clearly, if $u_2(x,t=0)=v_1(x,t=0)=0$, then both fields will vanish at all times and consequently the back-reaction will be zero at all times. Similar considerations apply when $u_1(x,t=0)=v_2(x,t=0)=0$.

Suppose that we initialize the scalar field at a boosted kink-antikink configuration 
\begin{equation}
\chi(x,t)=\chi_k(x+x_0,t;v)-\chi_k(x-x_0,t;-v)-1.
\end{equation} 
Moreover, we consider a single fermion at the zero mode bound to the kink, $\psi(x,t)=\psi_0(x+x_0;v)$. This configuration obeys $u_2(x,t=0)=v_1(x,t=0)=0$ and the back-reaction vanishes at all times. In order to turn on the back-reaction, we can consider the symmetric configurations discussed in section \ref{sec_sym}. These are the attractive and repulsive cases described by eq.~(\ref{eq_z2}) with $A=\pm 1$, and the case with the excited fermion is described by eq.~(\ref{eq_e2}) with $A=+1$. The latter is also attractive.  

We integrate the equations of motion in a box in the interval $-100.0<x<100.0$ by discretizing space in intervals with width $\Delta x=0.05$. The partial derivatives with respect to $x$ were approximated using a five-point stencil. The resulting set of equations was integrated using a fifth-order Runge-Kutta method with adaptive step size. Moreover, we considered periodic boundary conditions. When $\phi_0$ is very small, the error in the energy conservation may increase due to the increased strength of the nonlinear back-reaction term. The error tolerance in energy conservation in our numerical simulation is below $10^{-4}$. However, for large $\phi_0$, the error is much less significant.

\subsection{Fermions at the zero mode}
\begin{figure}[tbp]
\centering
   \includegraphics[width=0.75\linewidth]{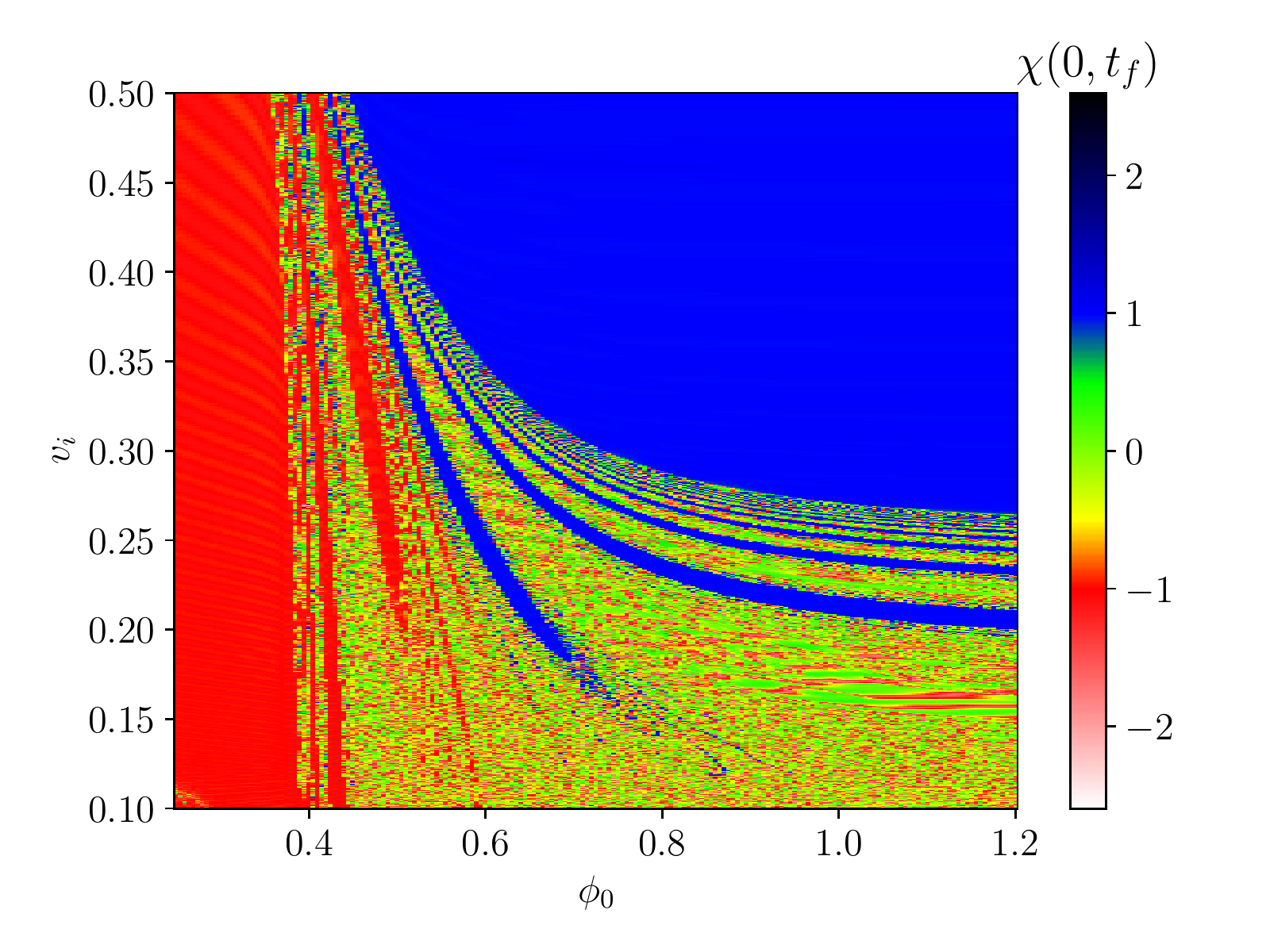}
   \caption{Final value of the scalar field at the center of the collision as a function of $\phi_0$ and $v$. We consider the attractive case where the fermions start at the zero mode.}
   \label{fig_mat}
\end{figure}

\begin{figure}[tbp]
\centering
   \includegraphics[width=0.75\linewidth]{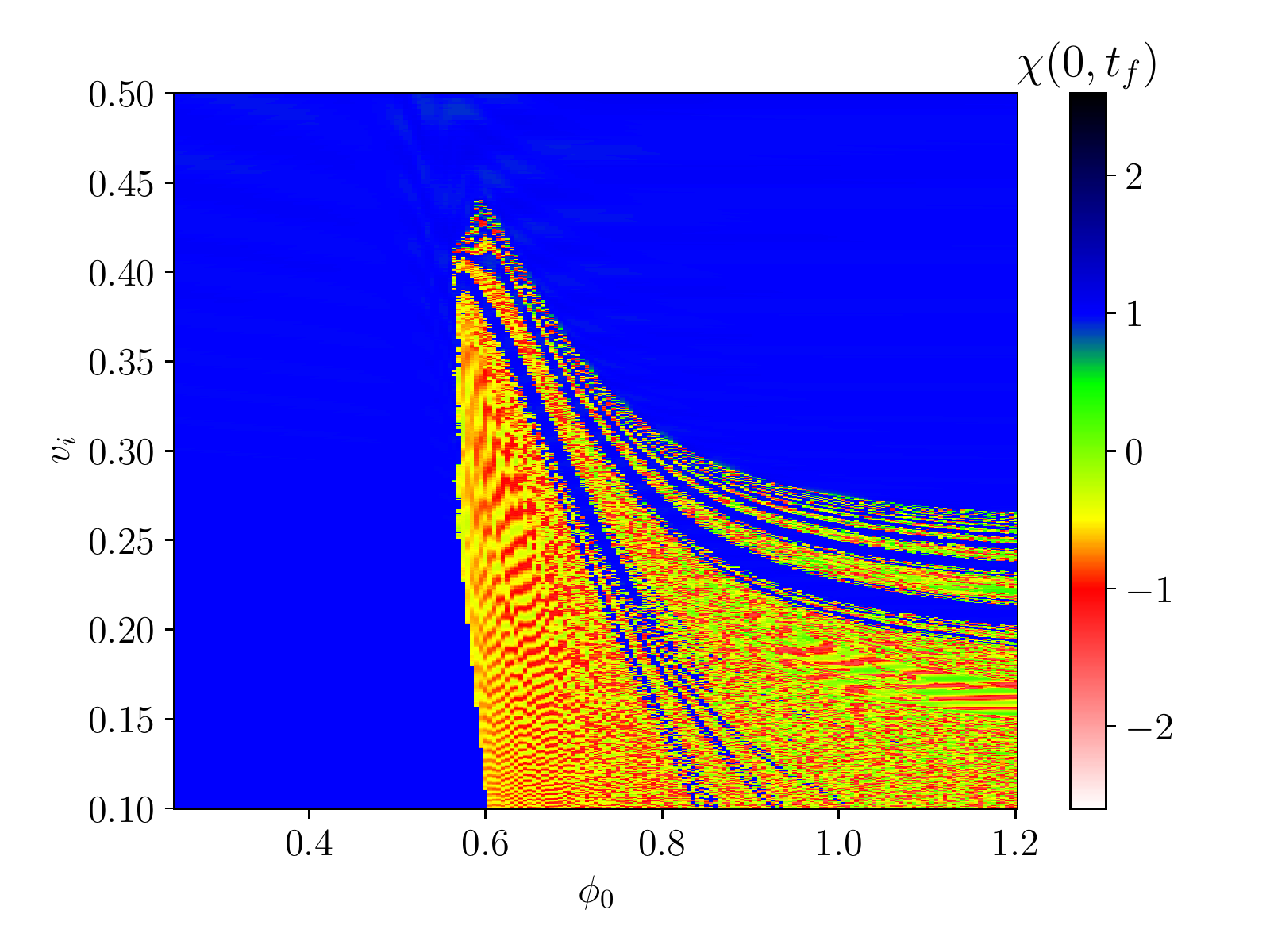}
   \caption{Final value of the scalar field at the center of the collision as a function of $\phi_0$ and $v$. We consider the repulsive case where the fermions start at the zero mode.}
   \label{fig_mat2}
\end{figure}

\begin{figure}[tbp]
\centering
   \begin{subfigure}[b]{1.0\textwidth}         
         \centering
         \includegraphics[width=\textwidth]{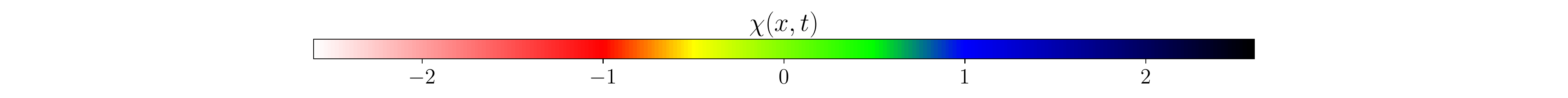}
   \end{subfigure}
   \begin{subfigure}[b]{0.32\textwidth}         
         \centering
         \includegraphics[width=\textwidth]{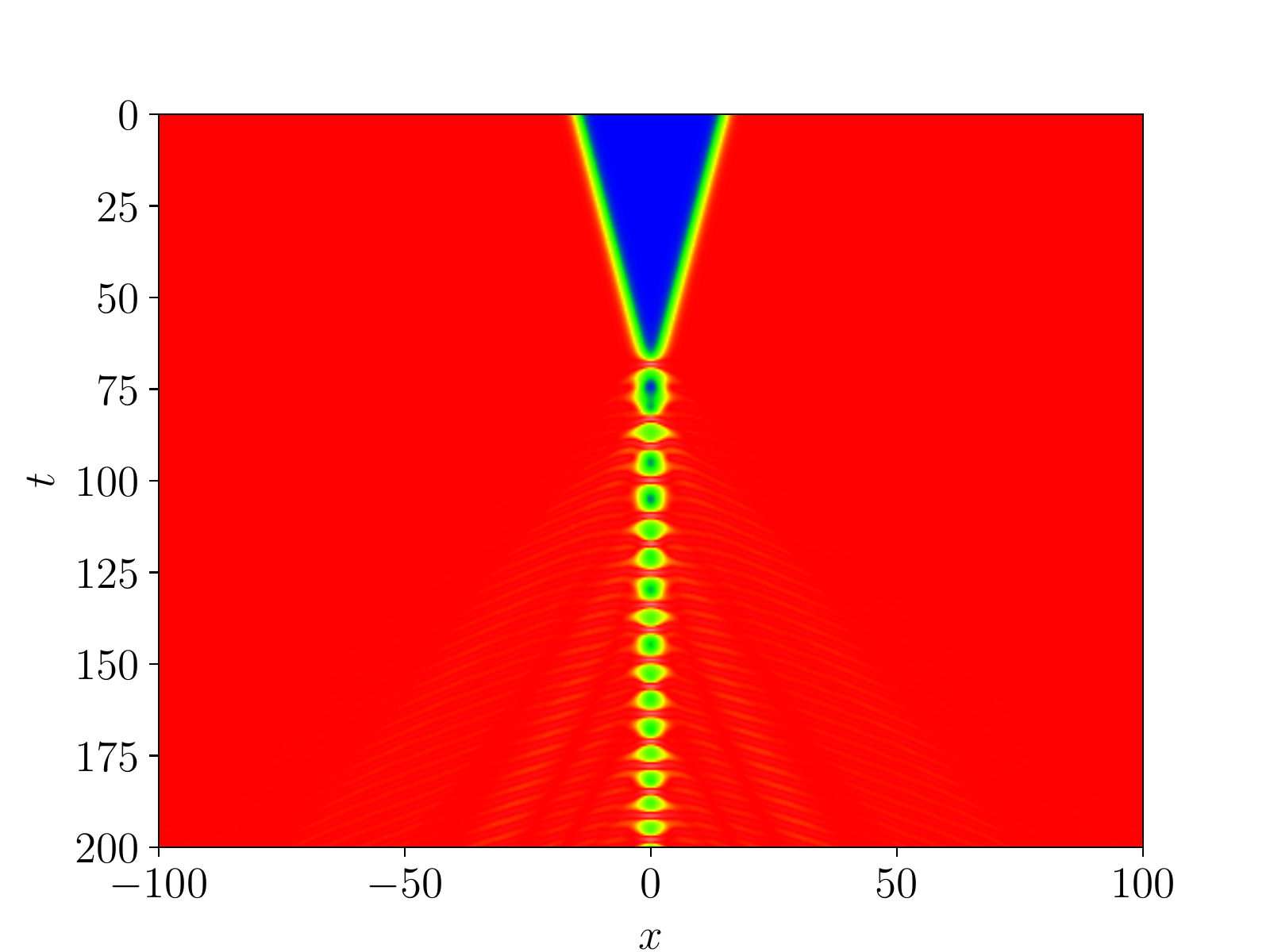}
         \caption{$\phi_0=0.8$, $v_i=0.2$}
     \end{subfigure}
     \begin{subfigure}[b]{0.32\textwidth}         
         \centering
         \includegraphics[width=\textwidth]{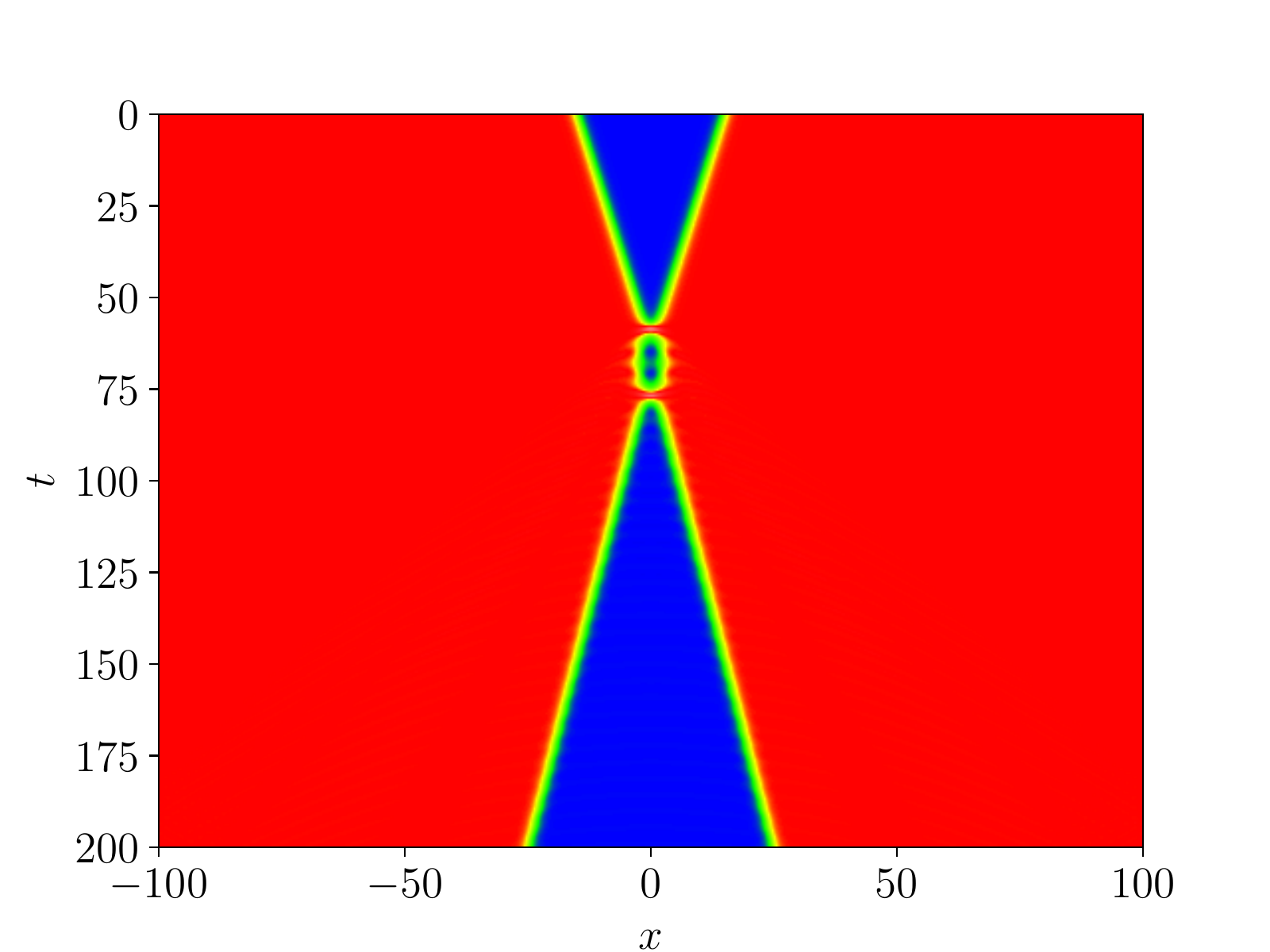}
         \caption{$\phi_0=0.8$, $v_i=0.235$}
     \end{subfigure}
     \begin{subfigure}[b]{0.32\textwidth}         
         \centering
         \includegraphics[width=\textwidth]{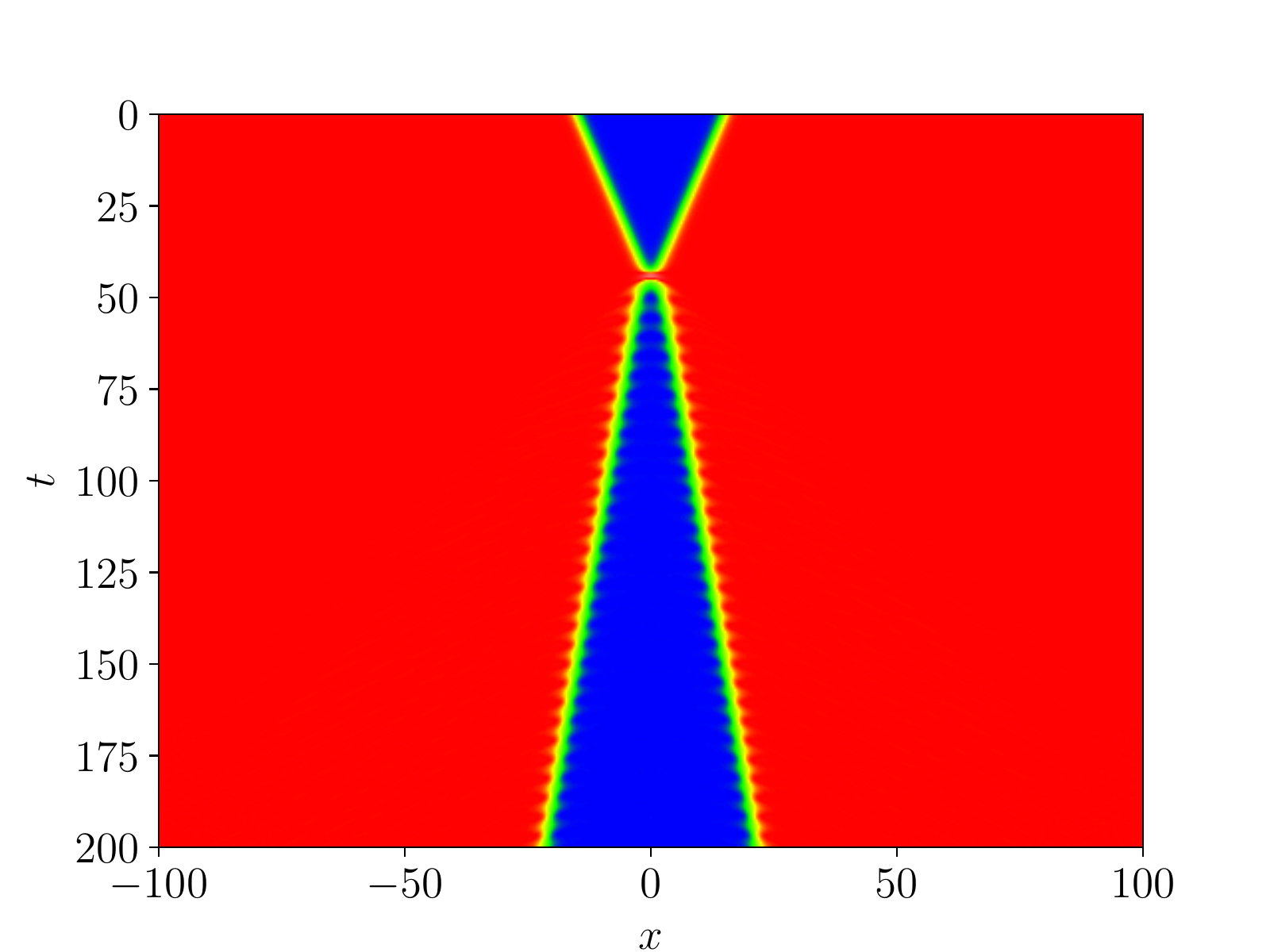}
         \caption{$\phi_0=0.8$, $v_i=0.32$}
     \end{subfigure}
     \begin{subfigure}[b]{0.32\textwidth}         
         \centering
         \includegraphics[width=\textwidth]{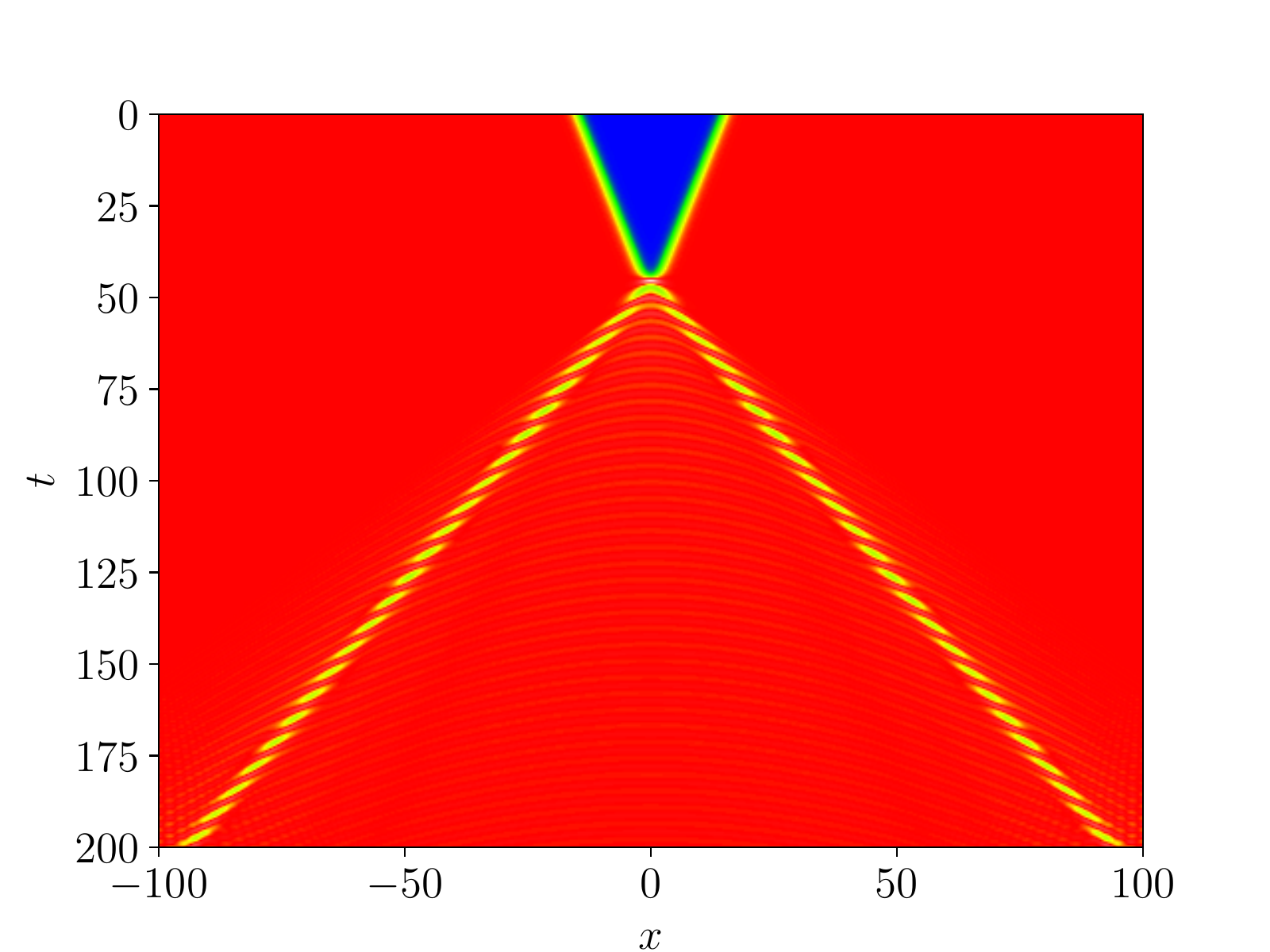}
         \caption{$\phi_0=0.3$, $v_i=0.3$}
     \end{subfigure}
     \begin{subfigure}[b]{0.32\textwidth}         
         \centering
         \includegraphics[width=\textwidth]{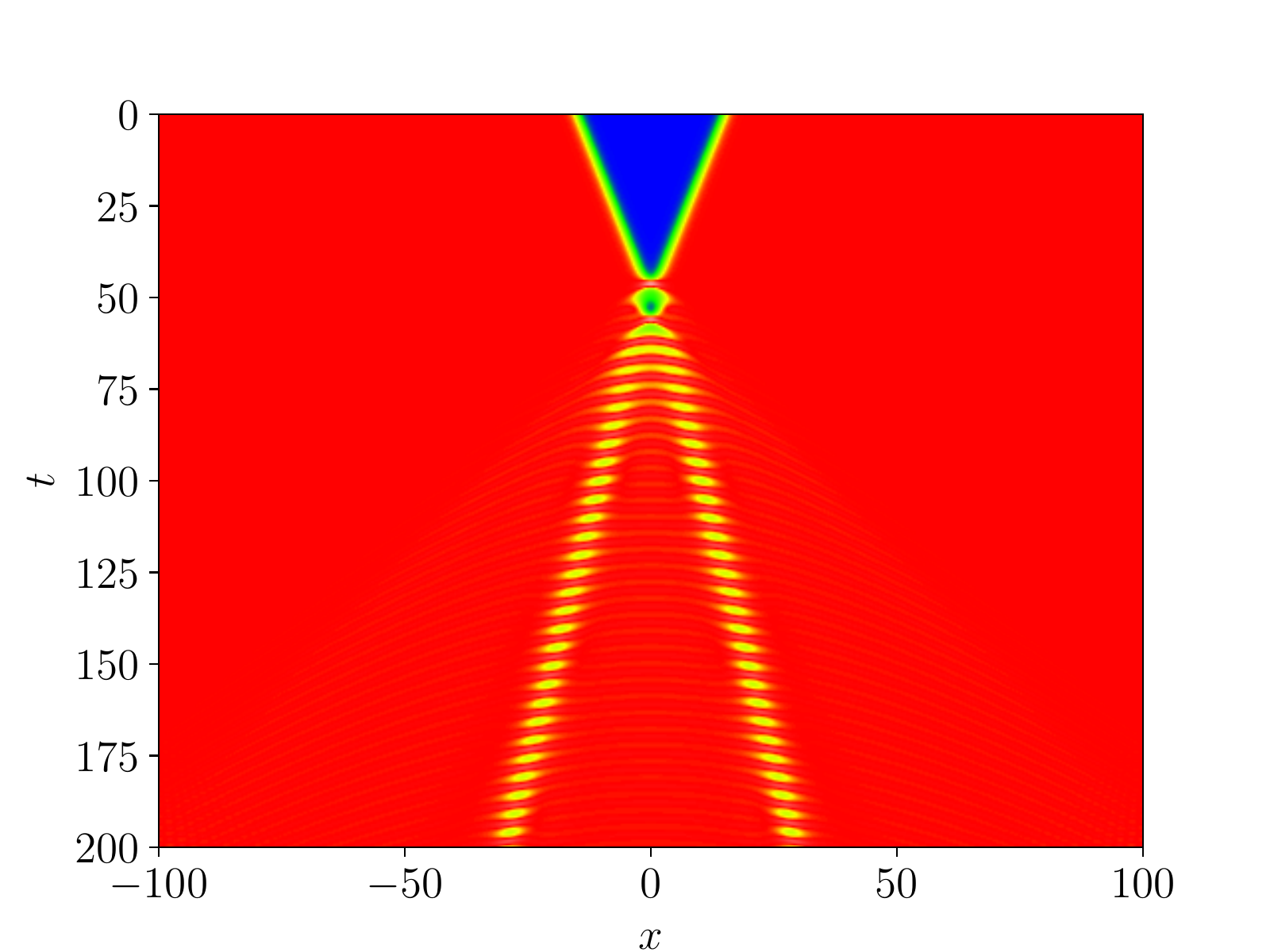}
         \caption{$\phi_0=0.475$, $v_i=0.3$}
     \end{subfigure}
     \begin{subfigure}[b]{0.32\textwidth}         
         \centering
         \includegraphics[width=\textwidth]{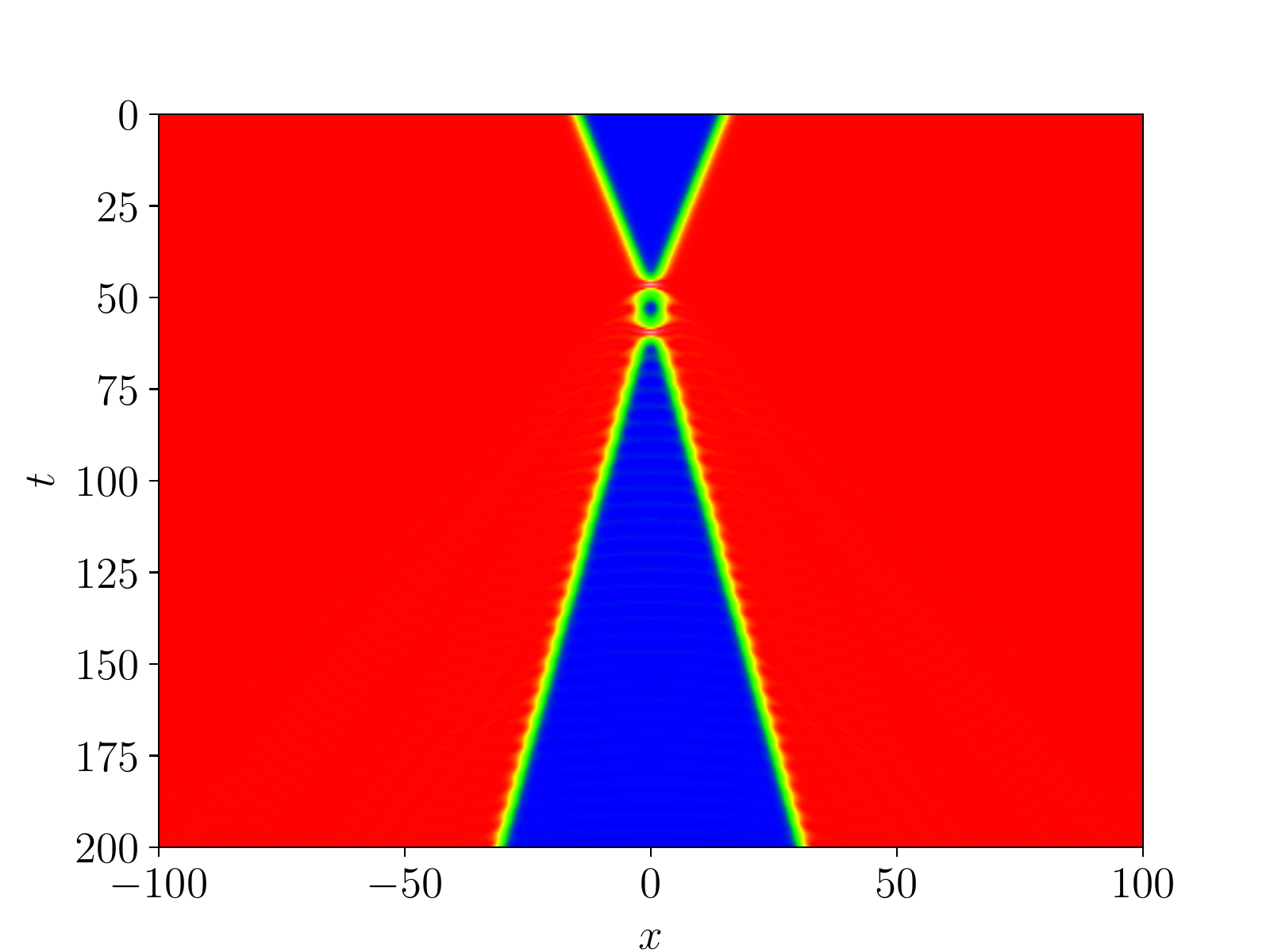}
         \caption{$\phi_0=0.545$, $v_i=0.3$}
     \end{subfigure}
     \caption{Scalar field evolution in spacetime. We consider the attractive case where the fermions start at the zero mode.}
   \label{fig_att}
\end{figure}

\begin{figure}[tbp]
\centering
     \begin{subfigure}[b]{1.0\textwidth}         
         \centering
         \includegraphics[width=\textwidth]{colorbar.pdf}
     \end{subfigure}
     \begin{subfigure}[b]{0.32\textwidth}         
         \centering
         \includegraphics[width=\textwidth]{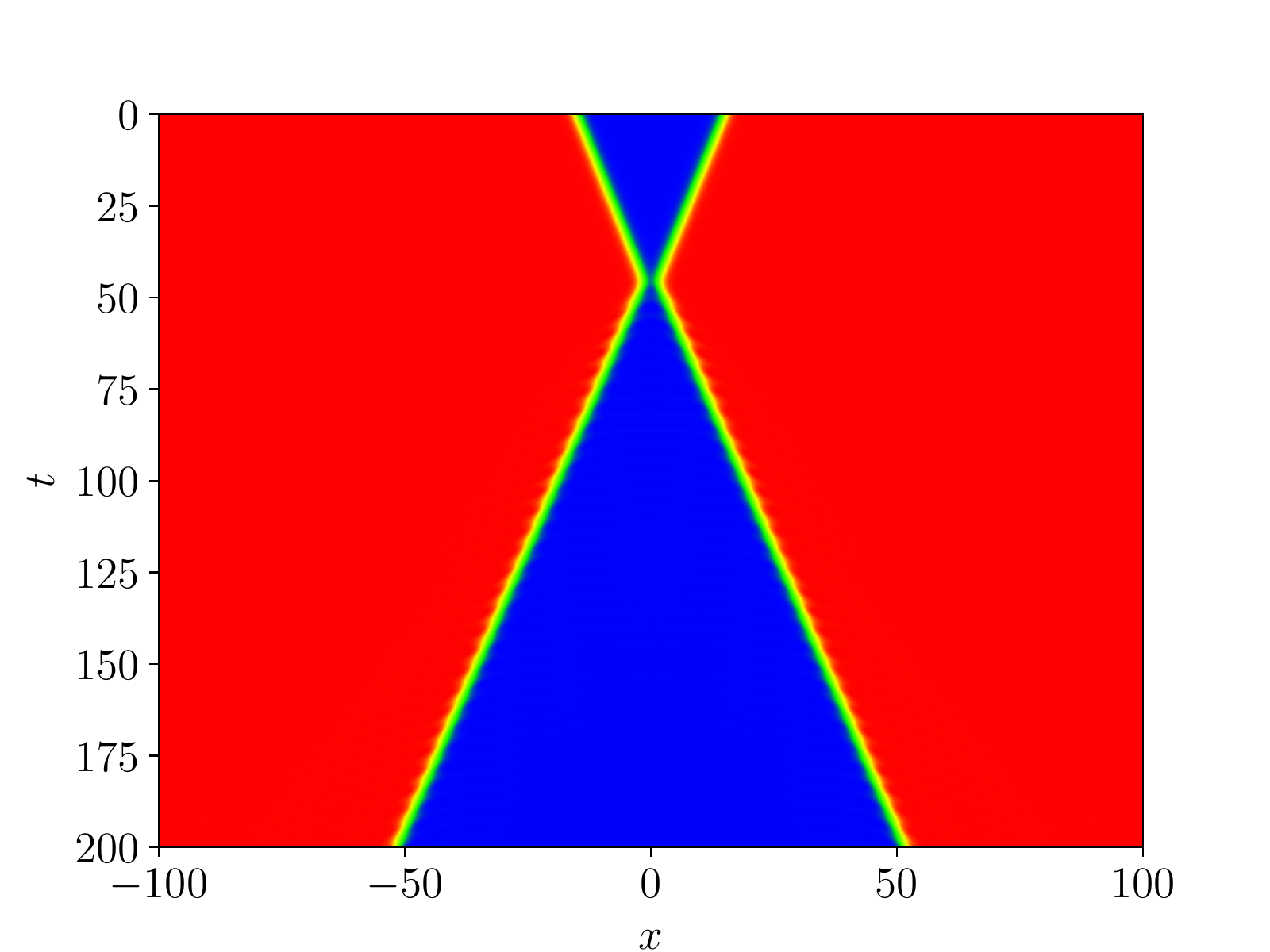}
         \caption{$\phi_0=0.4$, $v_i=0.3$}
     \end{subfigure}
     \begin{subfigure}[b]{0.32\textwidth}         
         \centering
         \includegraphics[width=\textwidth]{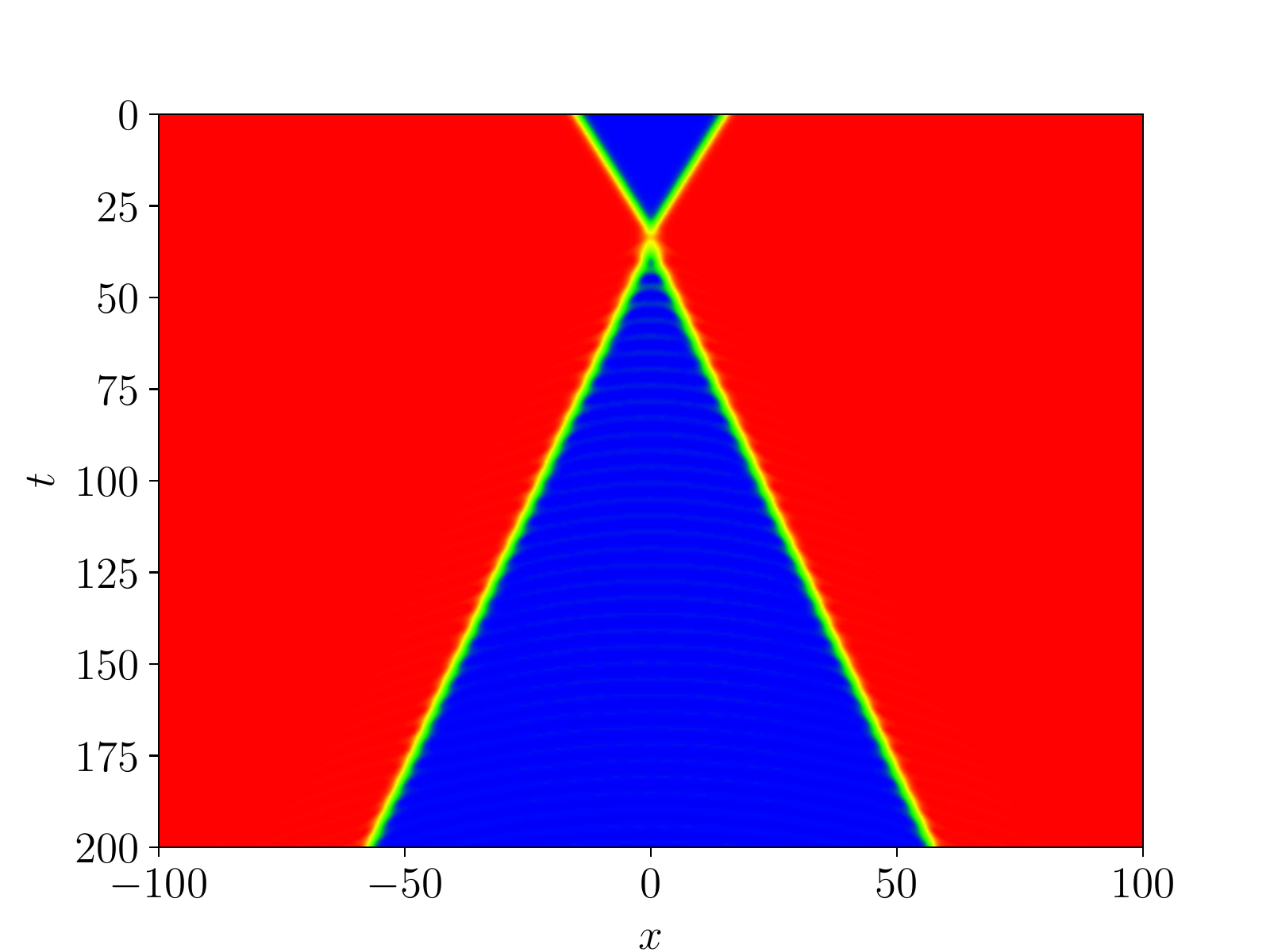}
         \caption{$\phi_0=0.54$, $v_i=0.47$}
     \end{subfigure}
     \begin{subfigure}[b]{0.32\textwidth}         
         \centering
         \includegraphics[width=\textwidth]{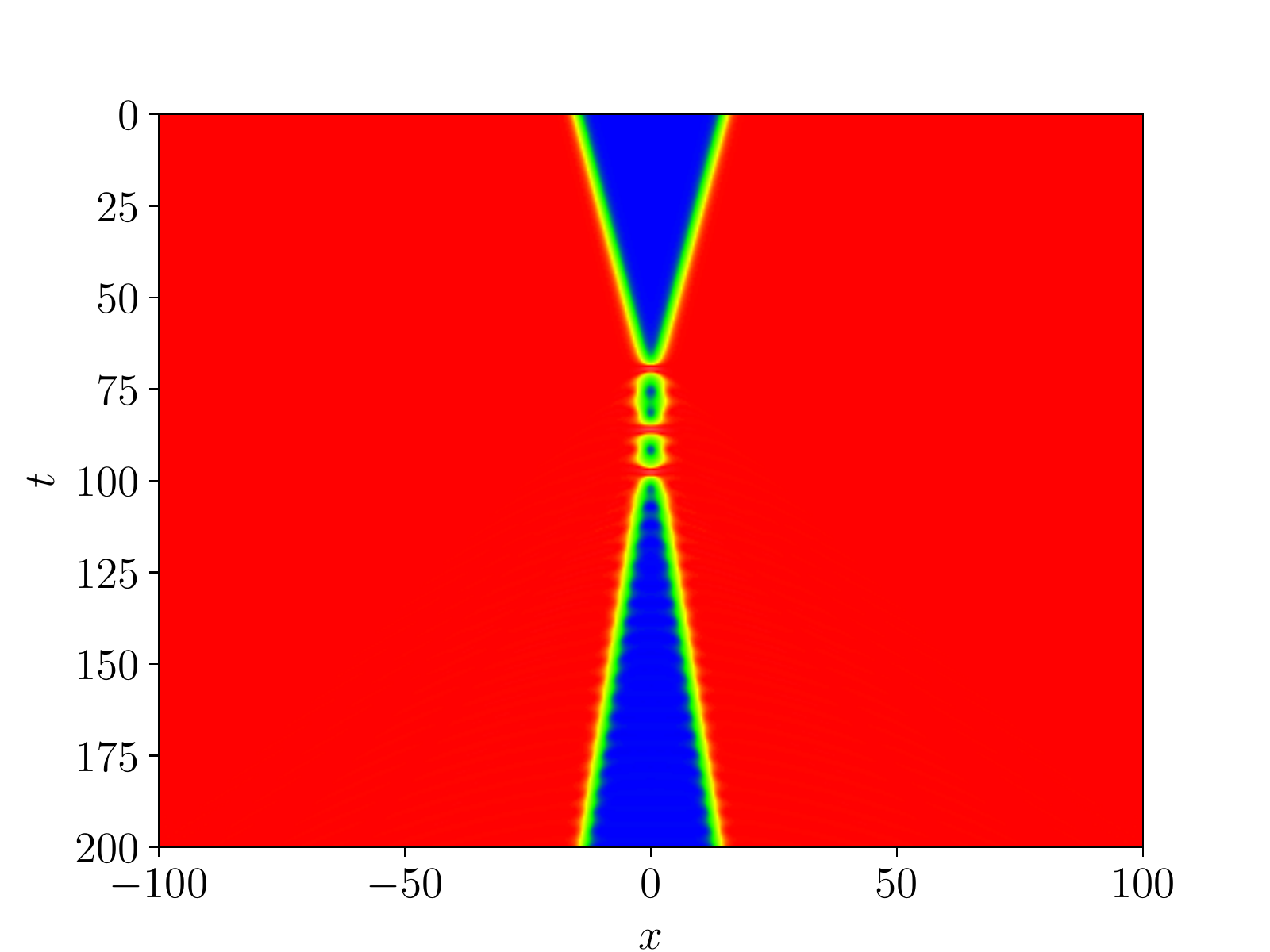}
         \caption{$\phi_0=1.083$, $v_i=0.2005$}
     \end{subfigure}
   \caption{Scalar field evolution in spacetime. We consider the repulsive case where the fermions start at the zero mode.}
   \label{fig_rep}
\end{figure}

In this section, we consider the configuration described by eqs.~(\ref{eq_z1}) and (\ref{eq_z2}).
The results for the attractive and repulsive cases are summarized in Figs.~\ref{fig_mat} and \ref{fig_mat2}, respectively. The colors represent the value of the scalar field at the origin and $t=40/v_i$. This value is plotted as a function of the velocity $v_i$ and the parameter $\phi_0$. Particular cases of the field's evolution in spacetime are shown in Figs.~\ref{fig_att} and \ref{fig_rep}.

The kink and the antikink interact and separate in the blue region in Figs.~\ref{fig_mat} and \ref{fig_mat2}, leaving the vacuum $\phi\simeq+1.0$ at the center. This happens in three scenarios: resonance windows, reflection, and repulsion without contact. Resonance windows and reflection happen in both repulsive and attractive cases. Fig.~\ref{fig_att}(b) gives an example for two-bounce resonance windows for the attractive case. A three-bounce example for the repulsive case is shown in Fig.~\ref{fig_rep}(c). A reflection between the kink and the antikink is shown in Fig.~\ref{fig_att}(c) for the attractive case.

For $\phi_0=1.2$, the structure of two-bounce resonance windows is similar to that in the absence of fermions. This is true in both scenarios. As $\phi_0$ decreases and the back-reaction becomes more significant, the resonance structure is deformed. In both cases, the critical velocity increases, and the windows become thinner. Moreover, a new window appears approximately in the interval $0.7\lesssim\phi_0\lesssim0.8$ for the attractive case. For the repulsive case, it appears for a slightly larger $\phi_0$. This window has one less oscillation between bounces than the lowest two-bounce resonance windows in the original $\phi^4$ model. An example of the evolution of the field in such a window is shown in Fig.~\ref{fig_att}(f). There are thinner windows near the large one, corresponding to separation after three bounces. 

Reflection without contact between the kink and the antikink happens only in the repulsive case for $\phi_0\lesssim0.6$. The repulsive force is strong enough to cause the kinks to recede before they collide in this region. An example is shown in Fig.~\ref{fig_rep}(a). The region where this behavior occurs merges with the reflection region in the upper right corner of Fig.~\ref{fig_mat2}. In between, we have a very soft bounce which is enough to reflect the kinks due to the repulsion of the fermionic field. An example is shown in Fig.~\ref{fig_rep}(b).
The yellow and green regions correspond to bion formation. It happens in both scenarios. An example is shown in Fig.~\ref{fig_att}(a). Finally, the kinks tend to form two oscillons after the collision when $\phi_0$ decreases in the attractive case. It leaves a vacuum $\phi\simeq-1.0$ at the center and corresponds to the red regions in Fig.~\ref{fig_mat}. These regions are called pseudo-windows. There is a large pseudo-window on the left side of Fig.~\ref{fig_mat}. The behavior in this region is illustrated in Fig.~\ref{fig_att}(d). Moreover, there are also many thinner pseudo-windows. They correspond to the formation of oscillons after more than a single bounce. We illustrate this behavior in Fig.~\ref{fig_att}(e).

In both Figs. \ref{fig_mat} and \ref{fig_mat2}, one can see that the critical velocity increases as $\phi_0$ is lowered. This means that the kinks need higher translational energy to overcome the mutual attraction after the collision. There are two possible explanations for this result. The first is that radiative losses increase as the back-reaction becomes more significant. The second one is that fermions are excited and store part of the energy. Unfortunately, it is hard to isolate the amount of energy lost as radiation and tell which effect is more important.

\begin{figure}[tbp]
\centering
   \begin{subfigure}[b]{1.0\textwidth}         
         \centering
         \includegraphics[width=\textwidth]{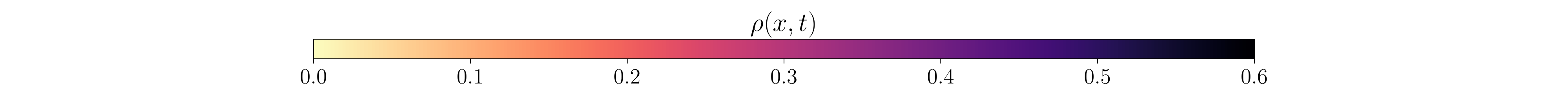}
   \end{subfigure}
   \begin{subfigure}[b]{0.32\textwidth}         
         \centering
         \includegraphics[width=\textwidth]{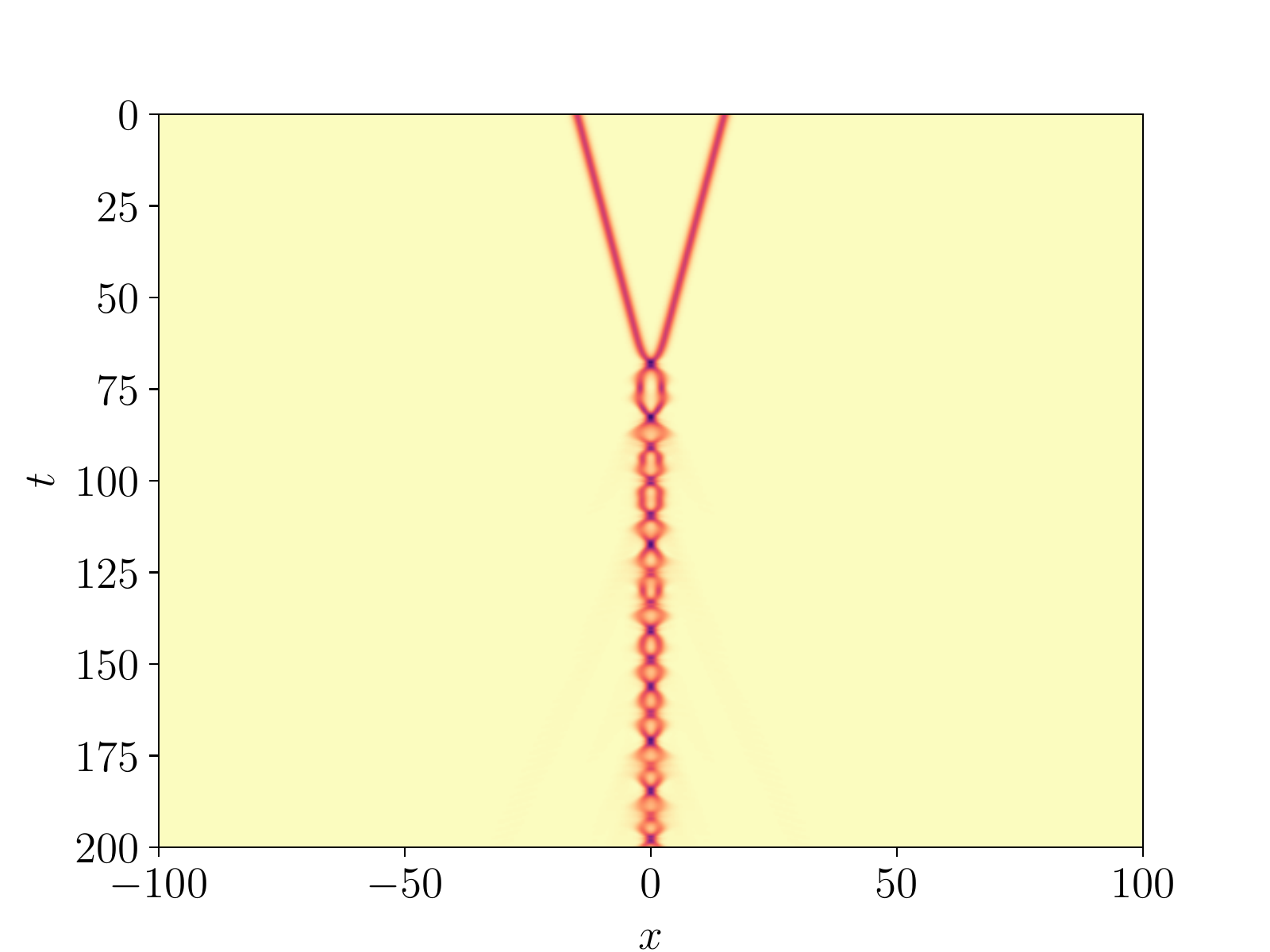}
         \caption{$\phi_0=0.8$, $v_i=0.2$}
     \end{subfigure}
     \begin{subfigure}[b]{0.32\textwidth}         
         \centering
         \includegraphics[width=\textwidth]{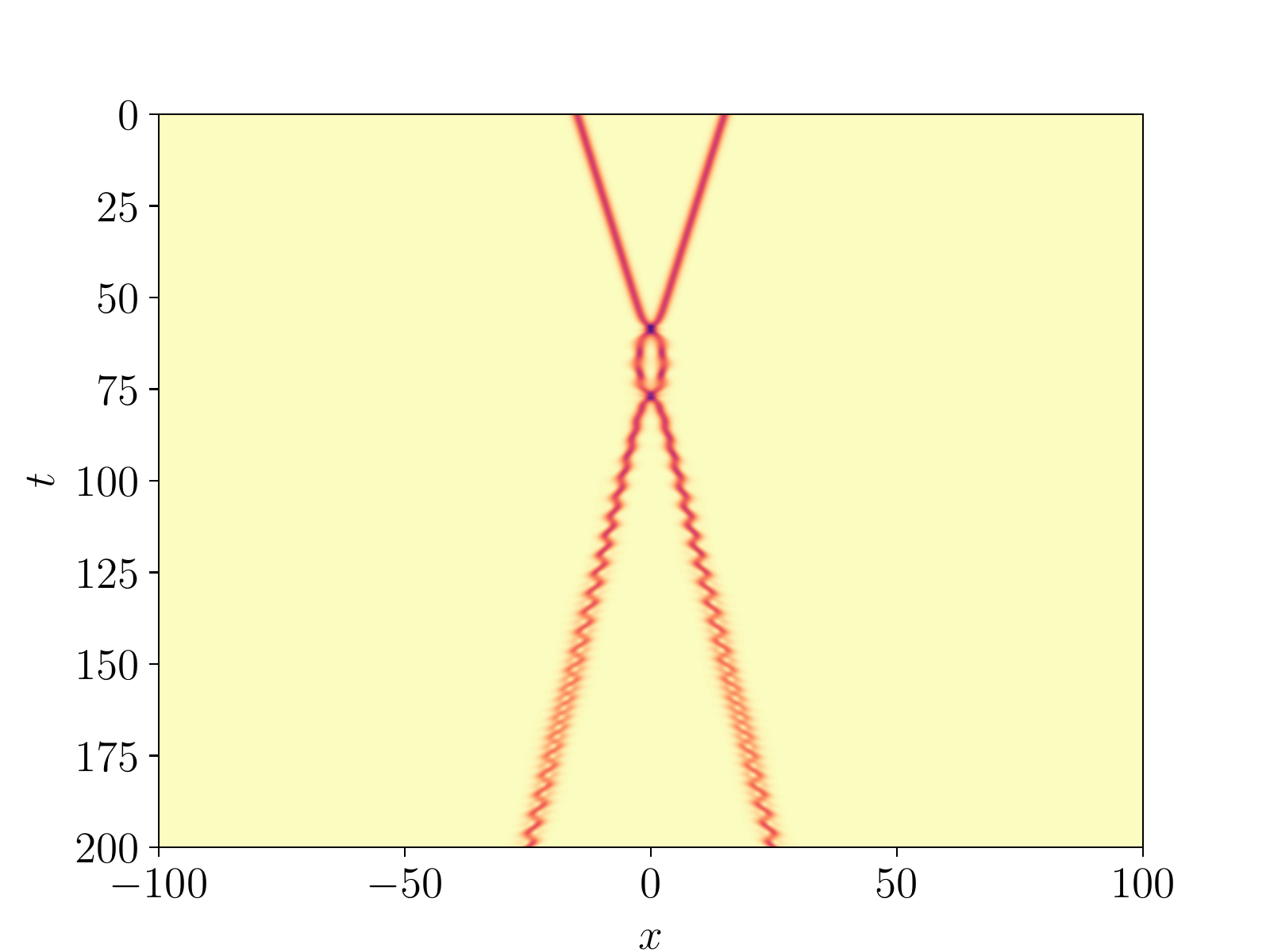}
         \caption{$\phi_0=0.8$, $v_i=0.235$}
     \end{subfigure}
     \begin{subfigure}[b]{0.32\textwidth}         
         \centering
         \includegraphics[width=\textwidth]{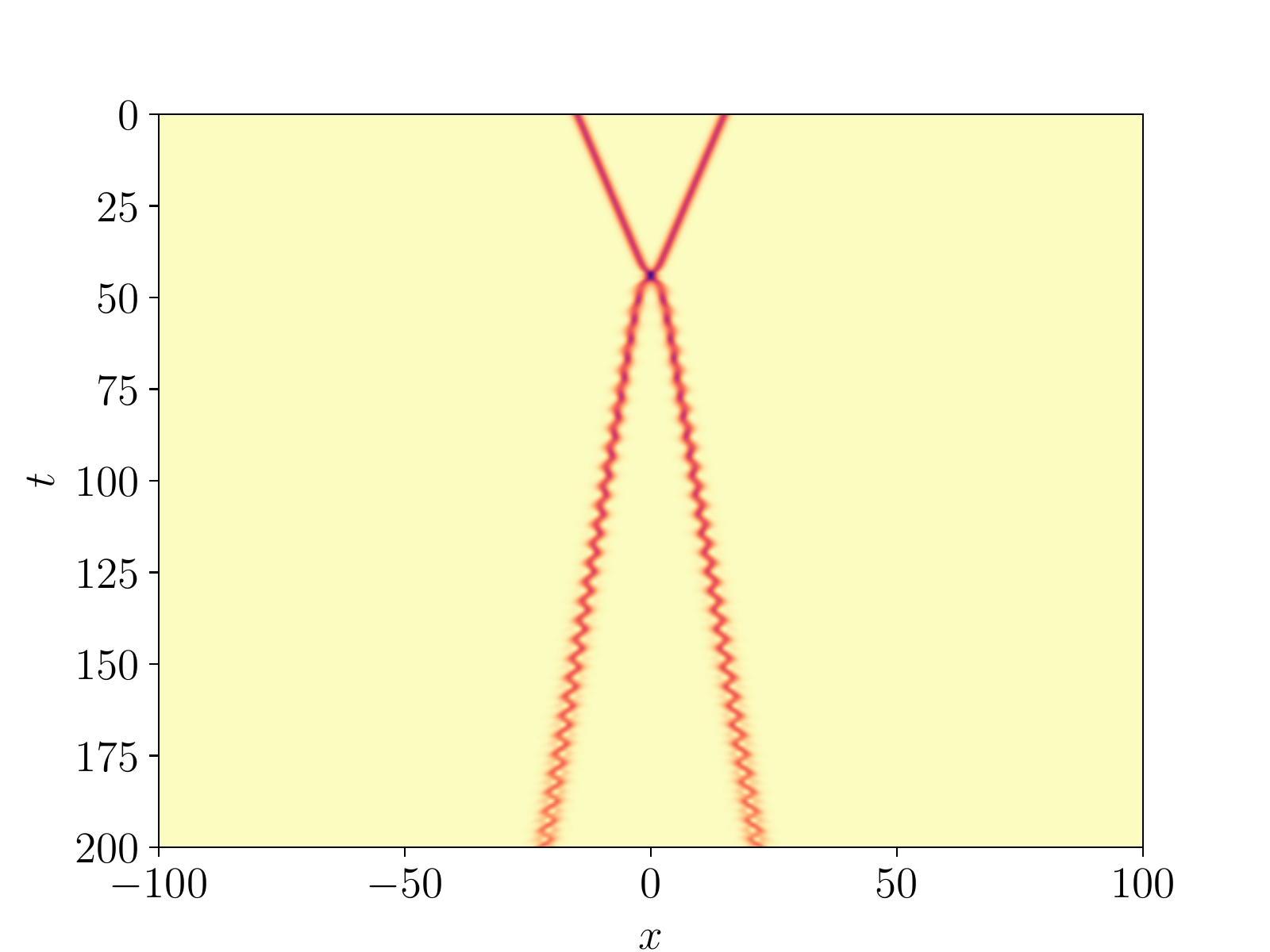}
         \caption{$\phi_0=0.8$, $v_i=0.32$}
     \end{subfigure}
     \begin{subfigure}[b]{0.32\textwidth}         
         \centering
         \includegraphics[width=\textwidth]{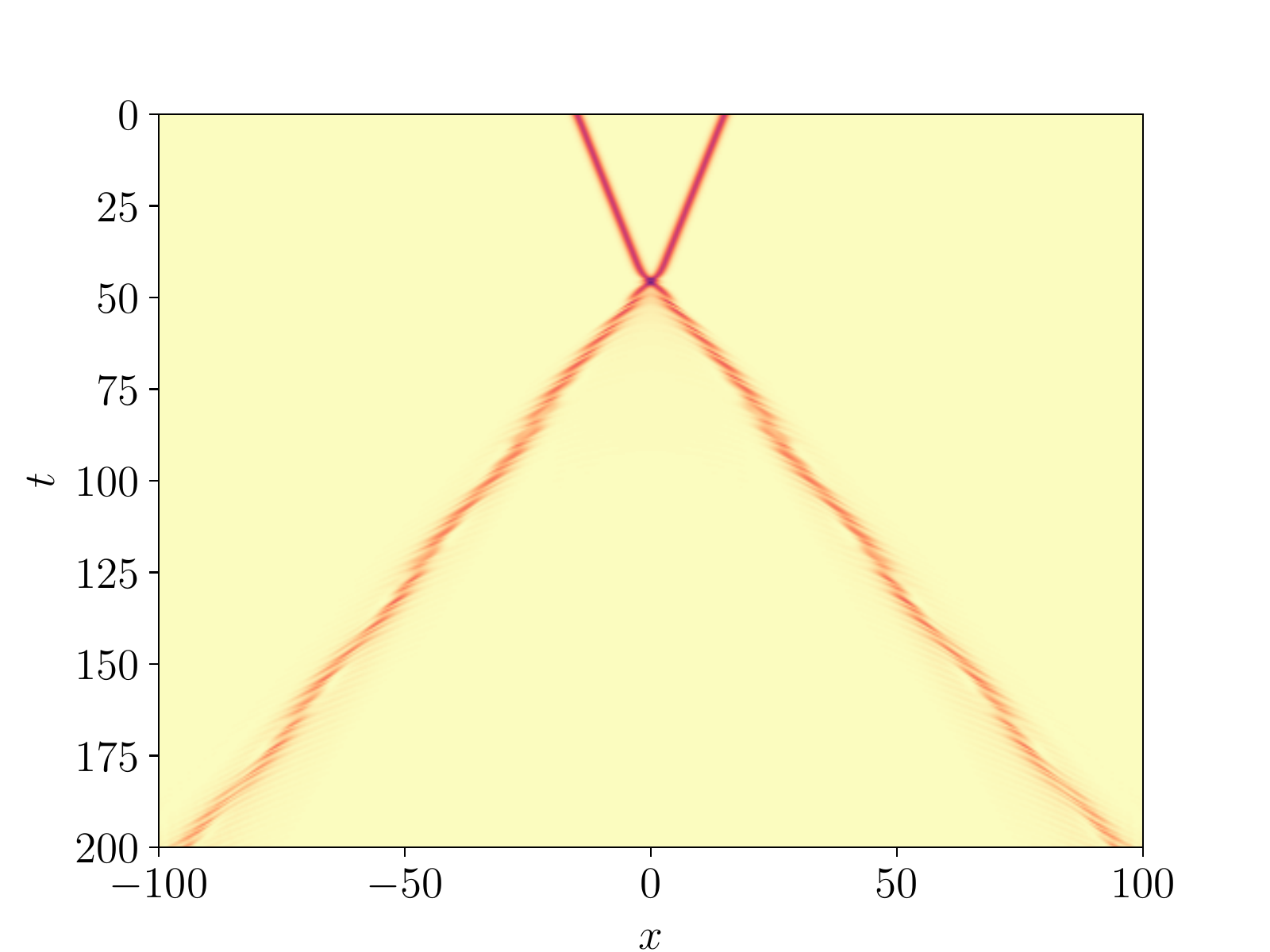}
         \caption{$\phi_0=0.3$, $v_i=0.3$}
     \end{subfigure}
     \begin{subfigure}[b]{0.32\textwidth}         
         \centering
         \includegraphics[width=\textwidth]{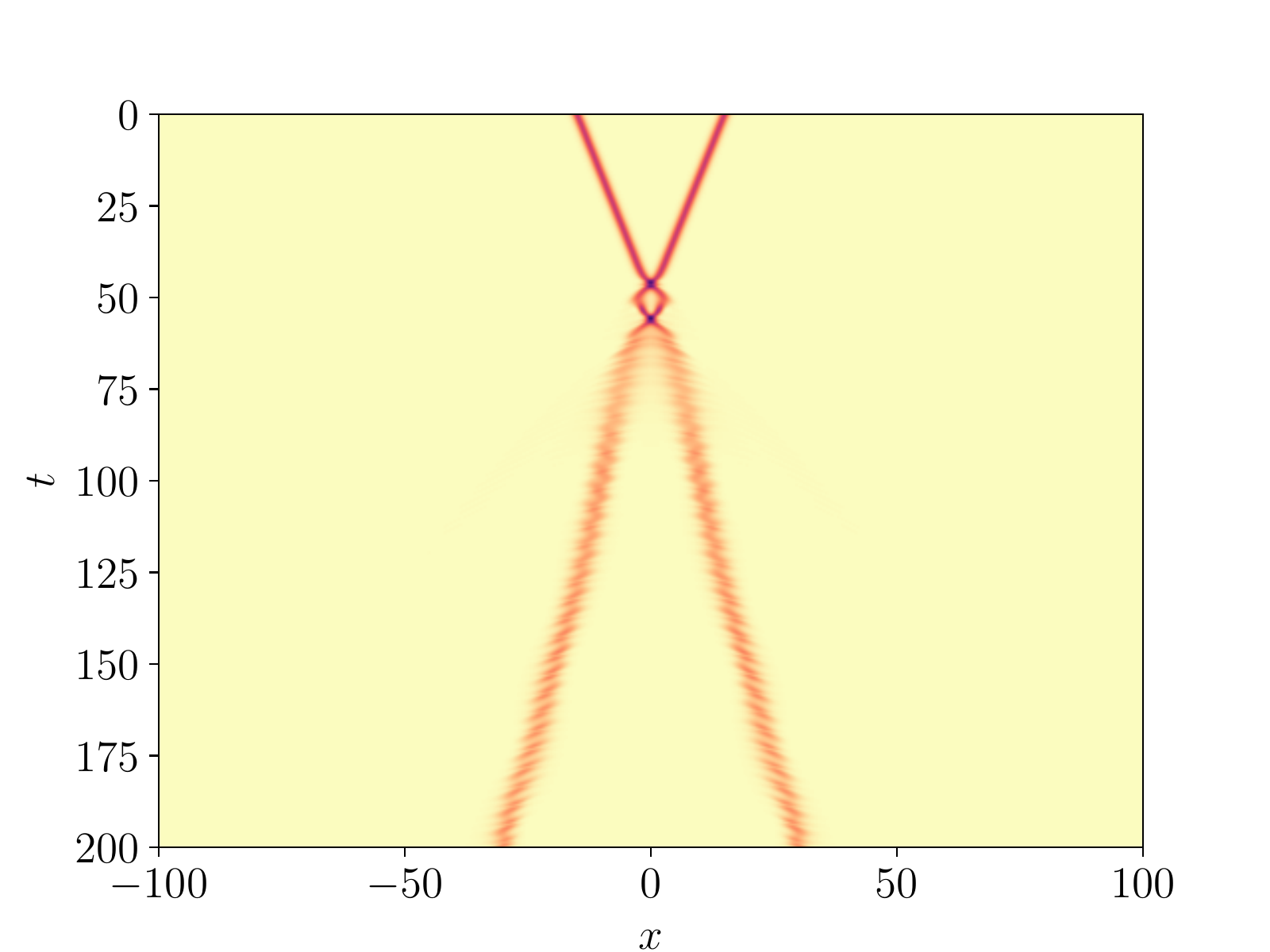}
         \caption{$\phi_0=0.475$, $v_i=0.3$}
     \end{subfigure}
     \begin{subfigure}[b]{0.32\textwidth}         
         \centering
         \includegraphics[width=\textwidth]{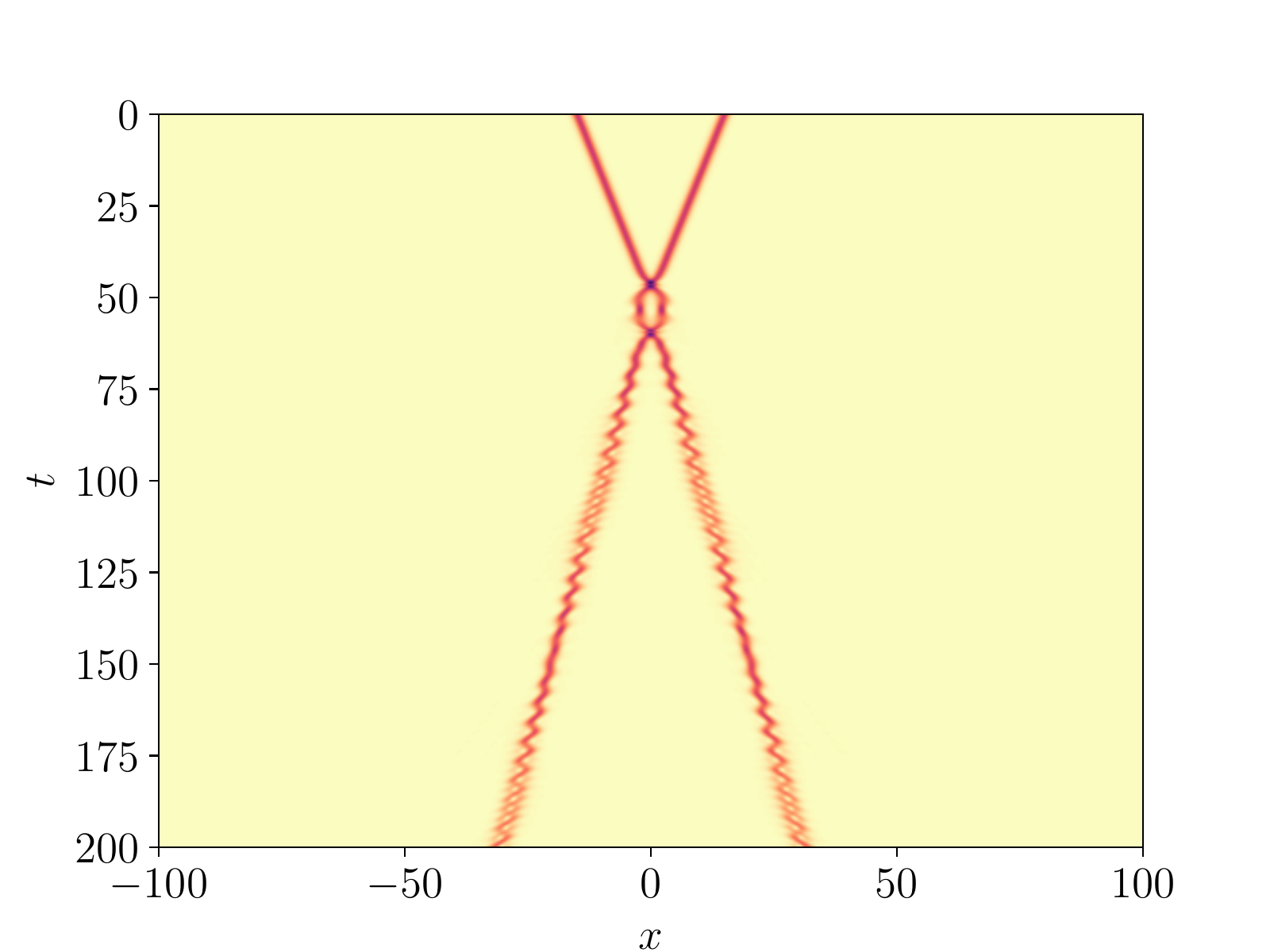}
         \caption{$\phi_0=0.545$, $v_i=0.3$}
     \end{subfigure}
     \caption{Evolution of the fermion density in spacetime. We consider the attractive case where the fermions start at the zero mode.}
   \label{fig_att_fermion}
\end{figure}

We also analyzed the evolution of the fermionic field during the collision. This is illustrated in Fig.~\ref{fig_att_fermion} for the same cases discussed in Fig.~\ref{fig_att}. In general, after the collision, the fermion stays bound to the kink, the bion, or the oscillon. Moreover, the fermion density exhibits oscillations afterward. They are expected for two reasons. First, we know that the scalar field internal modes are excited after the collision. Thus, the fermion interacts with an oscillating field \cite{campos2021fermions}. Second, fermion internal modes are also excited after the collision. However, the oscillations do not have a small amplitude and cannot be described as a linear combination of vibrational modes due to the highly nonlinear character of the equations of motion.

\begin{figure}[tbp]
\centering
     \begin{subfigure}[b]{1.0\textwidth}         
         \centering
         \includegraphics[width=\textwidth]{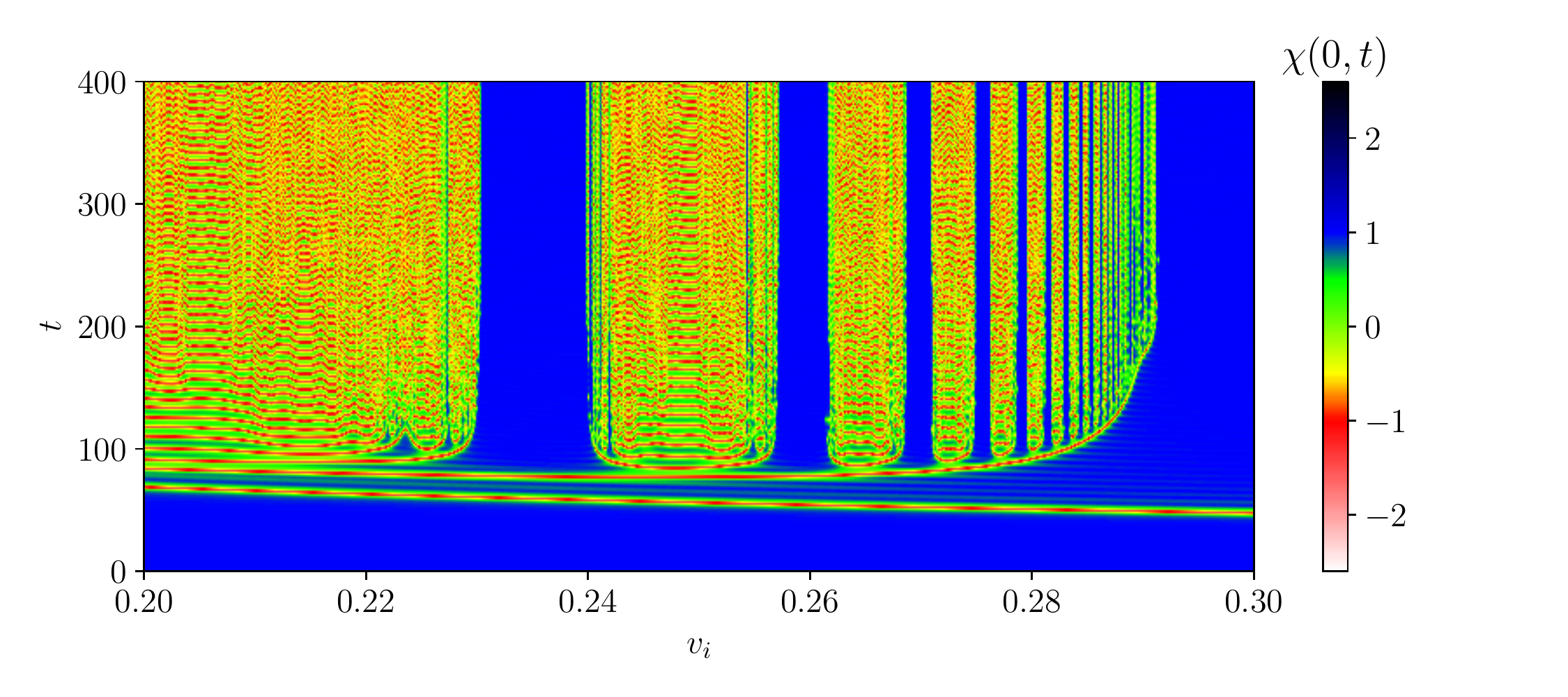}
         \caption{}
     \end{subfigure}
     \begin{subfigure}[b]{1.0\textwidth}         
         \centering
         \includegraphics[width=\textwidth]{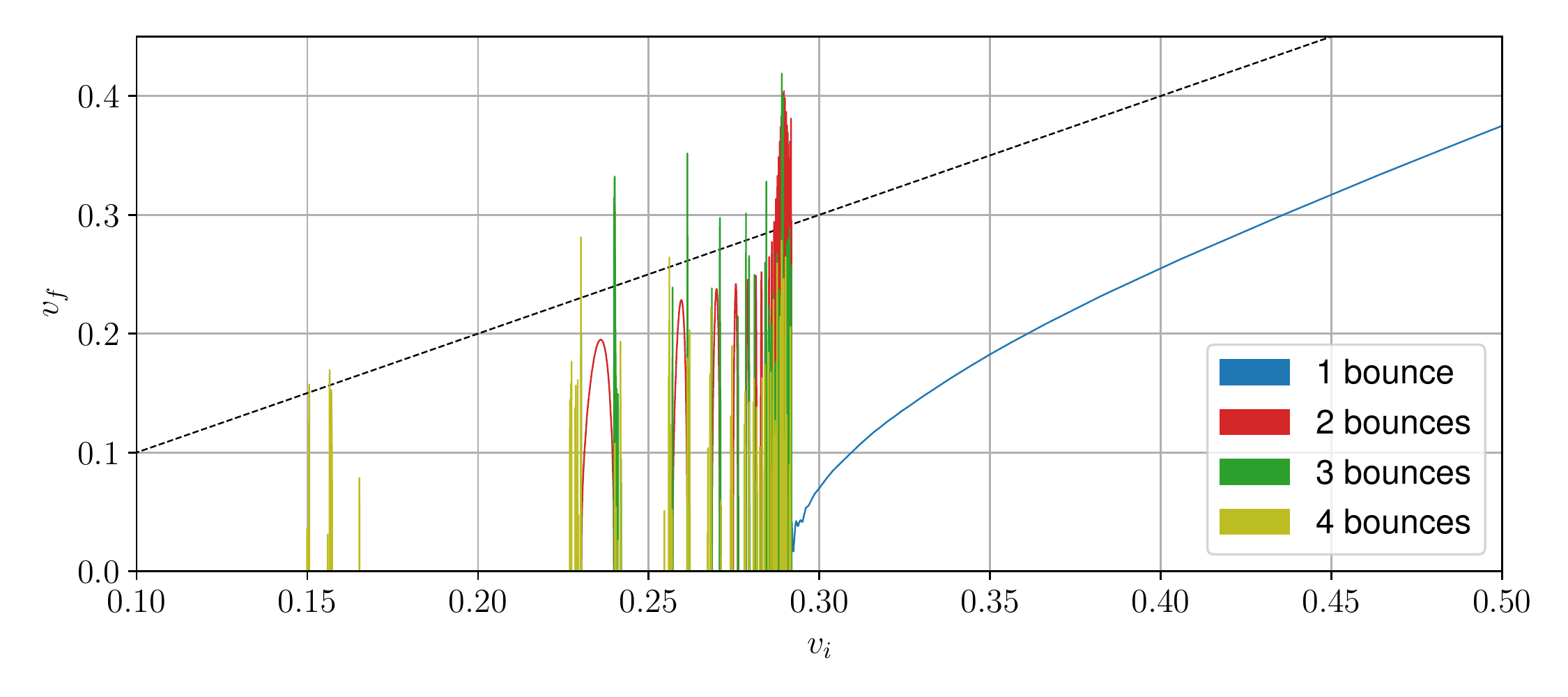}
         \caption{}
     \end{subfigure}
     \caption{(a) Field at the collision center as a function of time and initial velocity. (b) Kink's final velocity as a function of the initial one. Dashed line corresponds to the identity curve. We fix $\phi_0=0.8$.}
   \label{fig_res}
\end{figure}

The fractal character of bounce windows is an important property of resonant systems. To analyze further how this property is affected, we fix $\phi_0=0.8$. For this value, the back-reaction is already significant. Then, we measure the field at the collision center as a function of time and initial velocity. This is shown in Fig.~\ref{fig_res} (a). The instants where bounces occur can be easily spotted because they occur when the field crosses the value $-1$, marked by a red color. At bounce windows, the kinks separate, and the field at the center approaches the value $1$, marked in blue. Therefore, the resonance windows occur for values of $v_i$ where the field eventually reaches a constant blue color, creating a vertical stripe. In Fig.~\ref{fig_res} (b), we measure the final velocity $v_f$ as a function of the initial one. We searched for bounce windows in the range $0.1<v_i<0.5$ with velocity increments $\Delta v=10^{-5}$. We did the simulations up to a maximum time of $t=1400.0$. Whenever $v_f$ is non-vanishing, there is a bounce window. If $v_f$ is very small, the instant the separation occurs is large, and there is small imprecision in our measurement due to interaction with returning emitted radiation. The color indicates the number of bounces before separation. Interestingly, it is possible that $v_f>v_i$ in this model because the fermion can acquire negative energy and, thus, increase the kink's energy thanks to energy conservation. In both panels of Fig.~\ref{fig_res}, we see that the structure of higher bounce windows is still very rich. Therefore, the inclusion of back-reaction in our model does not seem to destroy the fractal property. However, as shown in Figs.~\ref{fig_mat} and \ref{fig_mat2}, if $\phi_0$ is decreased further, all resonance windows eventually disappear for both attractive and repulsive cases.

\begin{figure}[tbp]
\centering
   \begin{subfigure}[b]{0.48\textwidth}         
         \centering
         \caption{}         
         \includegraphics[width=\textwidth]{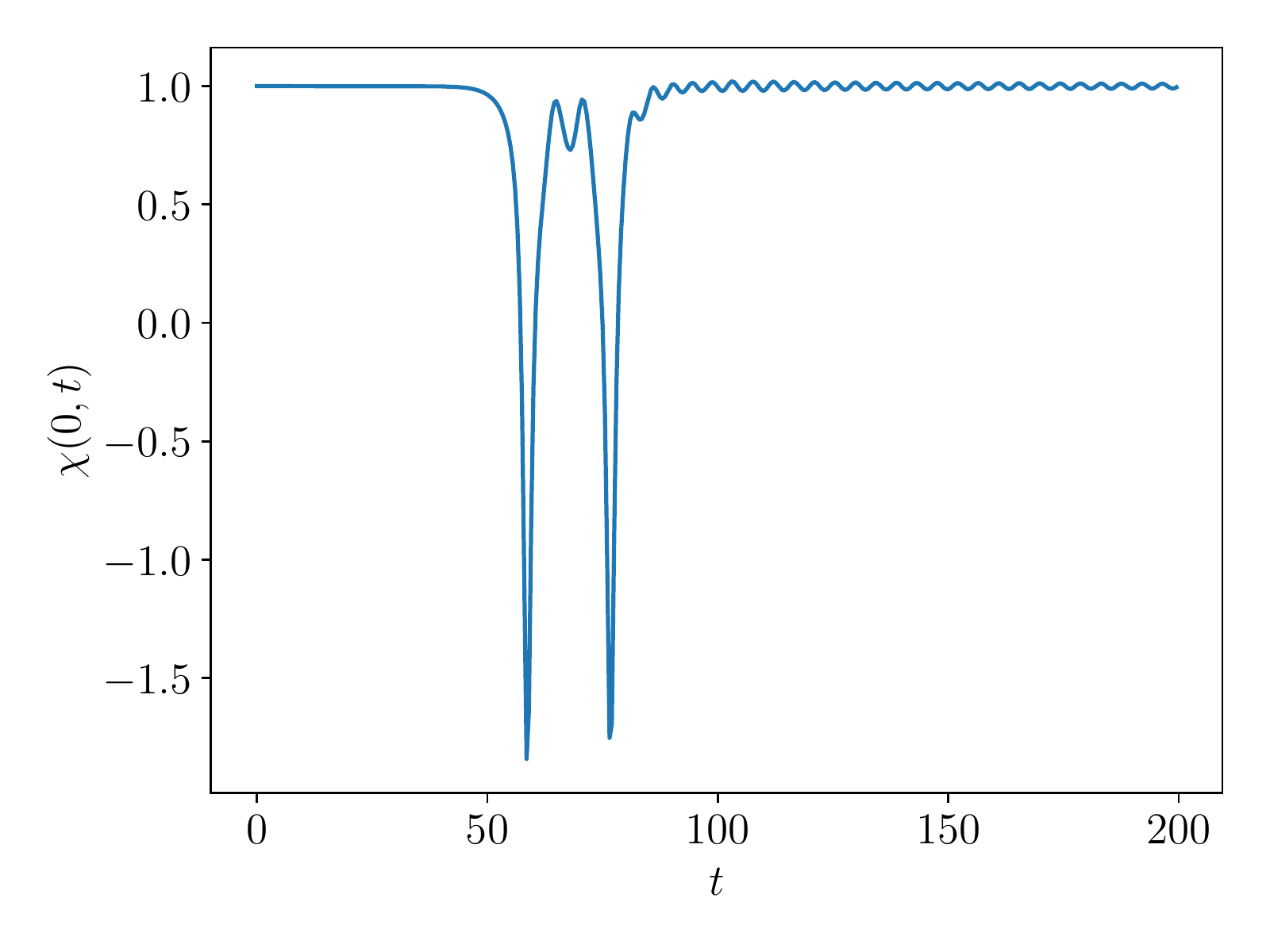}
         
     \end{subfigure}
     \begin{subfigure}[b]{0.48\textwidth}         
         \centering
         \caption{}         
         \includegraphics[width=\textwidth]{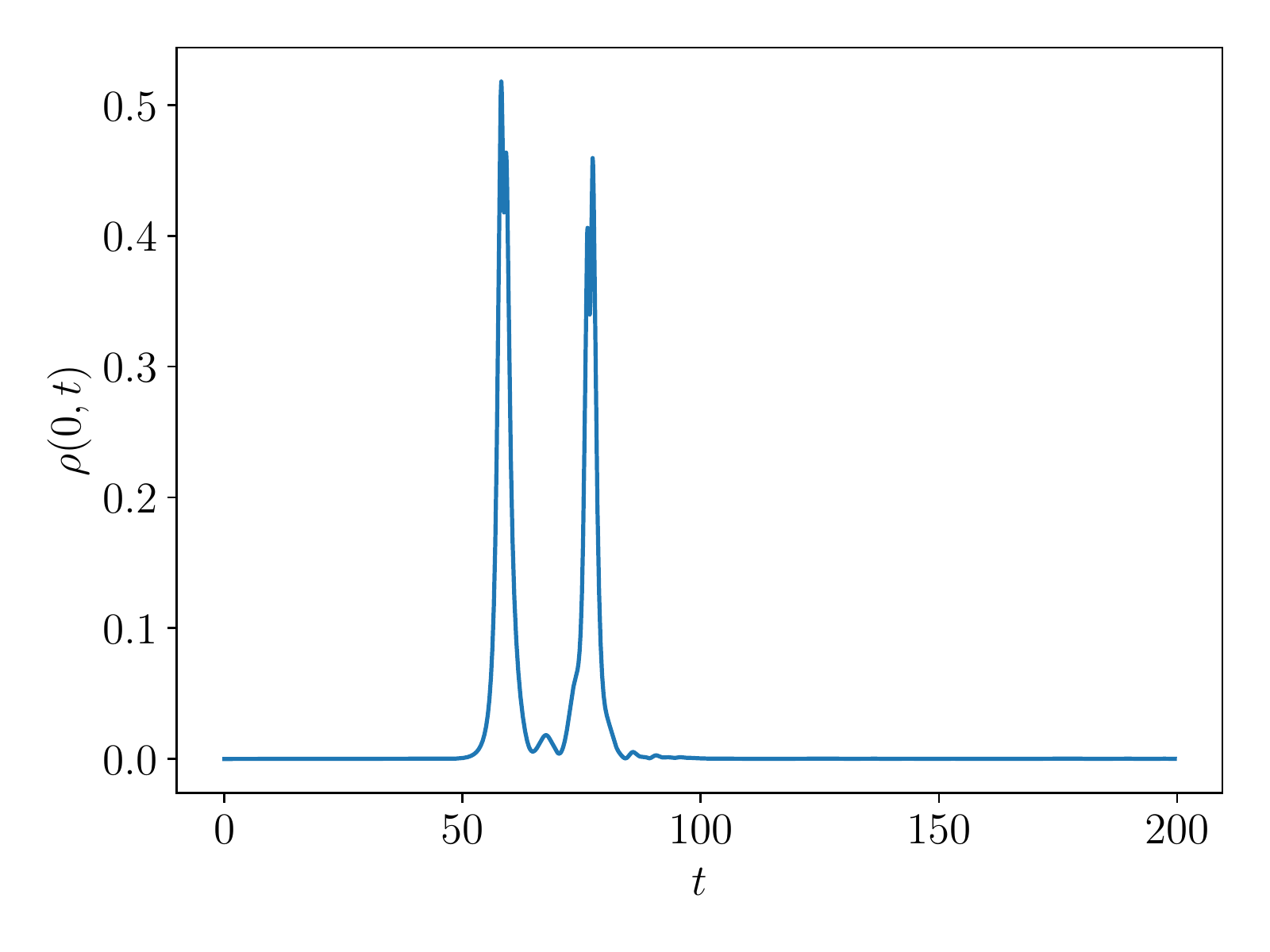}
     \end{subfigure}
     \caption{Evolution of (a) scalar field and (b) the fermion density at the center of the collision. We consider the attractive case where the fermions start at the zero mode. Parameters are $v_i=0.235$ and $\phi_0=0.8$.}
   \label{fig_att_cent}
\end{figure}

An interesting observation is that when we calculate the scalar field and the fermion density at the center of the collision, both have the same number of oscillations between bounces, as shown in Fig.~\ref{fig_att_cent}. This means that the fermion and kink oscillate at approximately the same frequency. This is expected when the fermions start at the zero-mode due to the degenerate excitation spectrum of the system, calculated in section \ref{sec_stab}.

\subsection{Fermions at the excited state}

\begin{figure}[tbp]
\centering
   \includegraphics[width=0.75\linewidth]{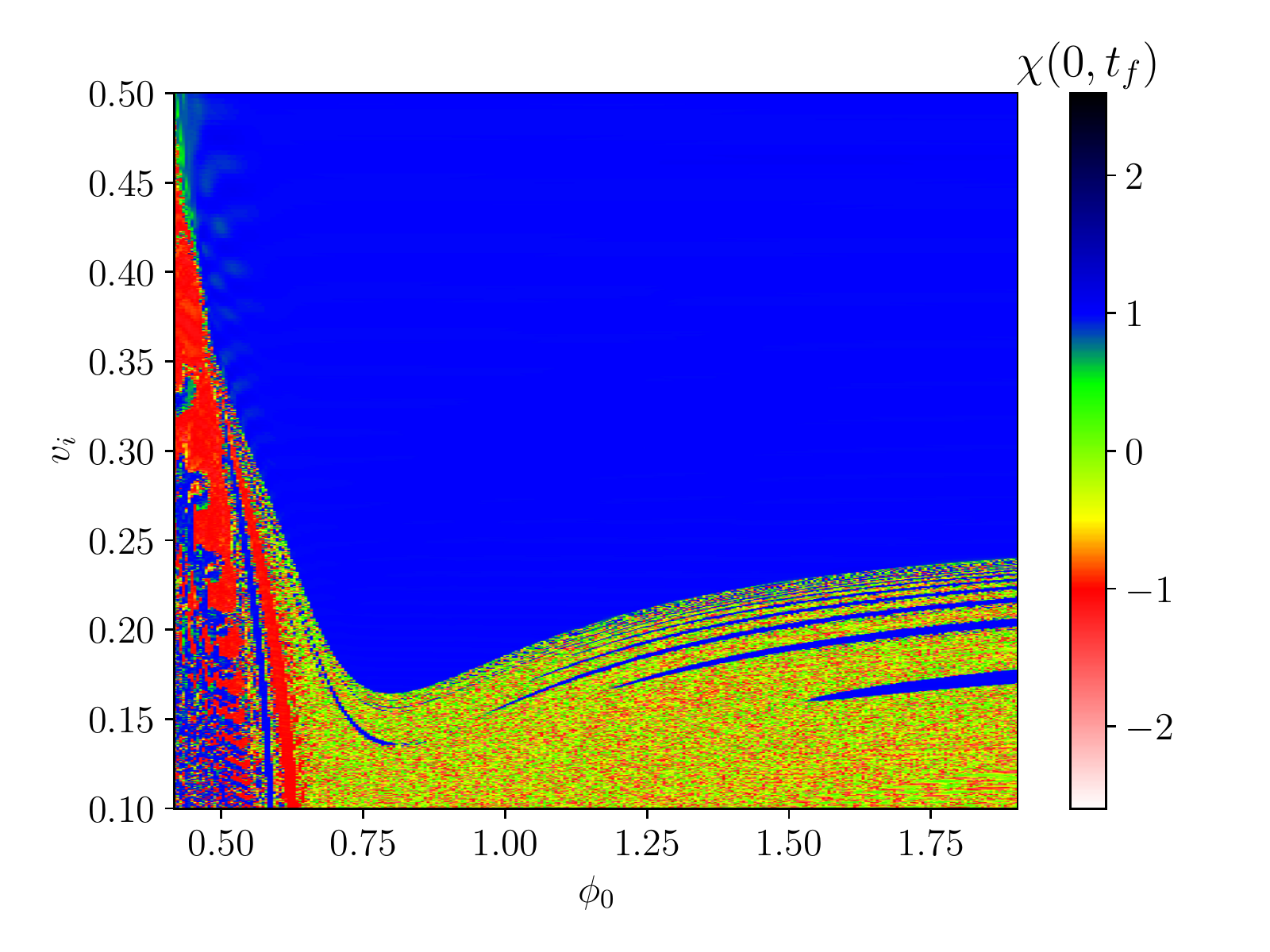}
   \caption{Final value of the scalar field at the center of the collision as a function of $\phi_0$ and $v_i$. We consider the case where the fermions start at the excited state.}
   \label{fig_mat3}
\end{figure}

Now, let us turn our attention to the case where the fermions are excited. There are many possibilities to explore in this case, but we will study only a single case for brevity. It is given by eq.~(\ref{eq_e2}) with $A=1$. The possible behaviors of the system are shown in Fig.~\ref{fig_mat3}. Once more, the colors represent the value of the scalar field at the origin and $t=40/v_i$. The blue region corresponds to reflection or resonance windows with one or multiple bounces. The yellow and green regions correspond to bion formation. Finally, the red region corresponds to pseudo-windows, where two oscillons or an extra kink-antikink pair are formed.

\begin{figure}[tbp]
\centering
   \begin{subfigure}[b]{1.0\textwidth}         
         \centering
         \includegraphics[width=\textwidth]{colorbar.pdf}
   \end{subfigure}
   \begin{subfigure}[b]{0.32\textwidth}         
         \centering
         \includegraphics[width=\textwidth]{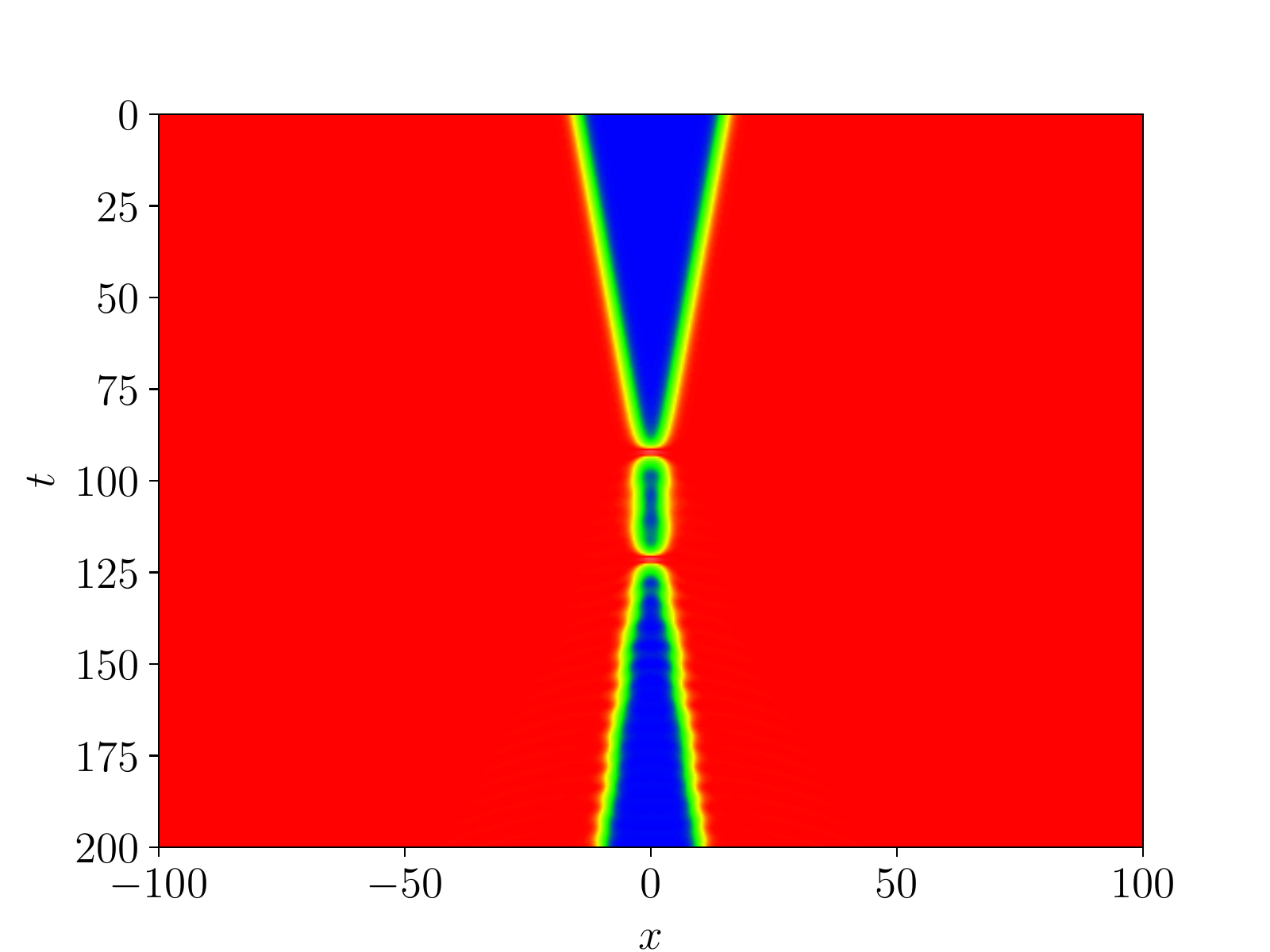}
         \caption{$\phi_0=0.76$, $v_i=0.14$}
     \end{subfigure}
     \begin{subfigure}[b]{0.32\textwidth}         
         \centering
         \includegraphics[width=\textwidth]{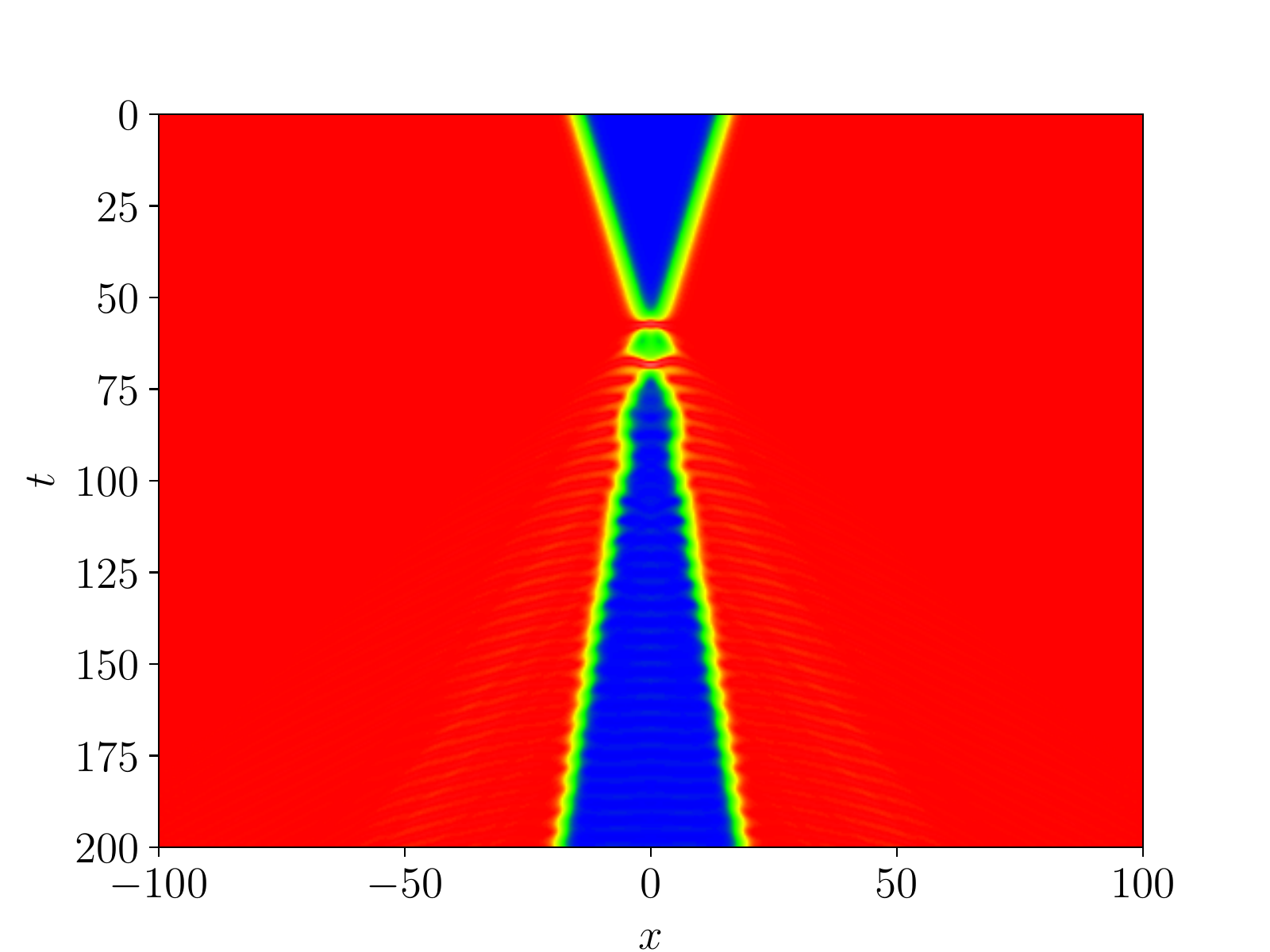}
         \caption{$\phi_0=0.55$, $v_i=0.23$}
     \end{subfigure}
     \begin{subfigure}[b]{0.32\textwidth}         
         \centering
         \includegraphics[width=\textwidth]{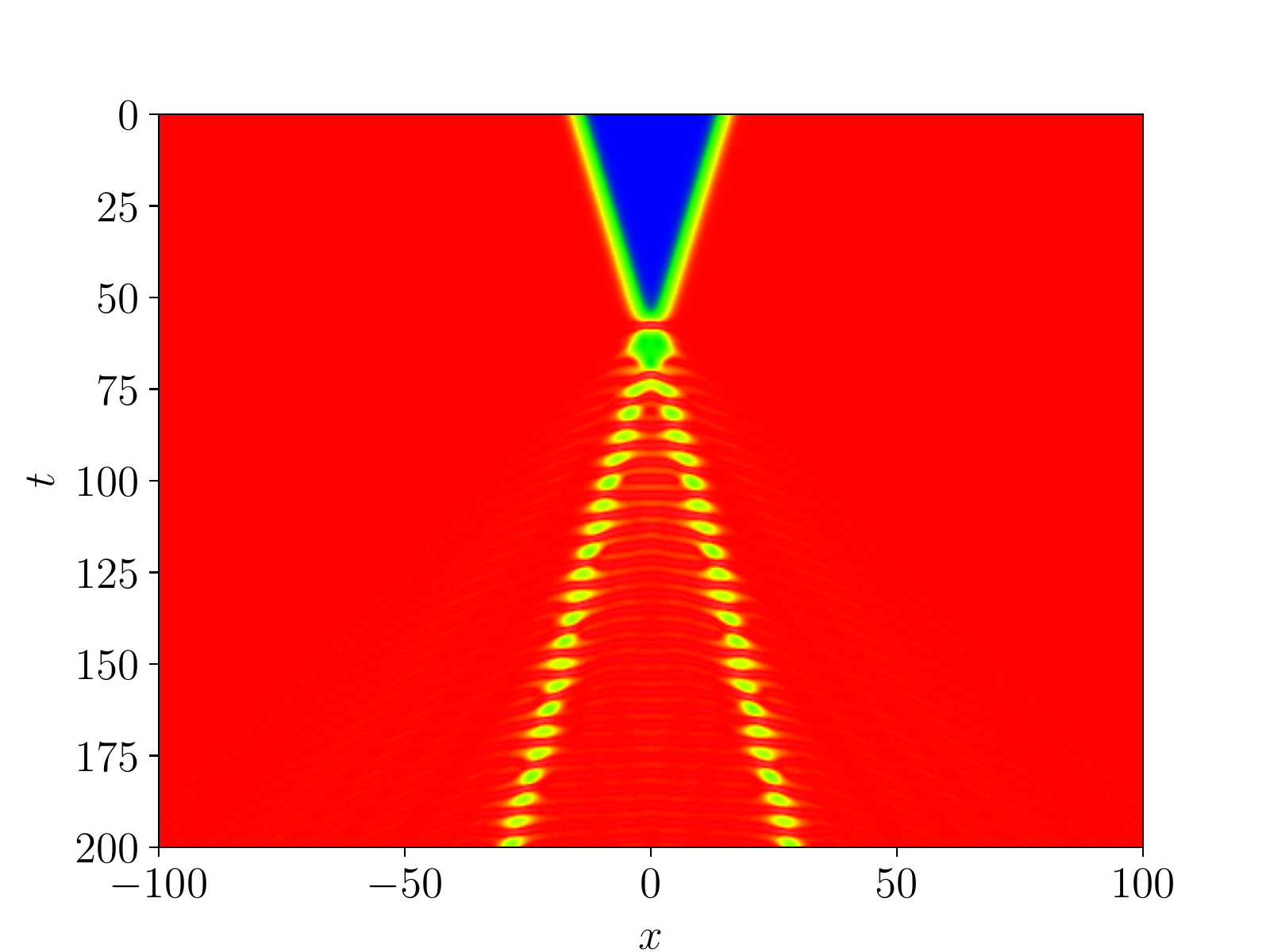}
         \caption{$\phi_0=0.575$, $v_i=0.23$}
     \end{subfigure}
     \begin{subfigure}[b]{0.32\textwidth}         
         \centering
         \includegraphics[width=\textwidth]{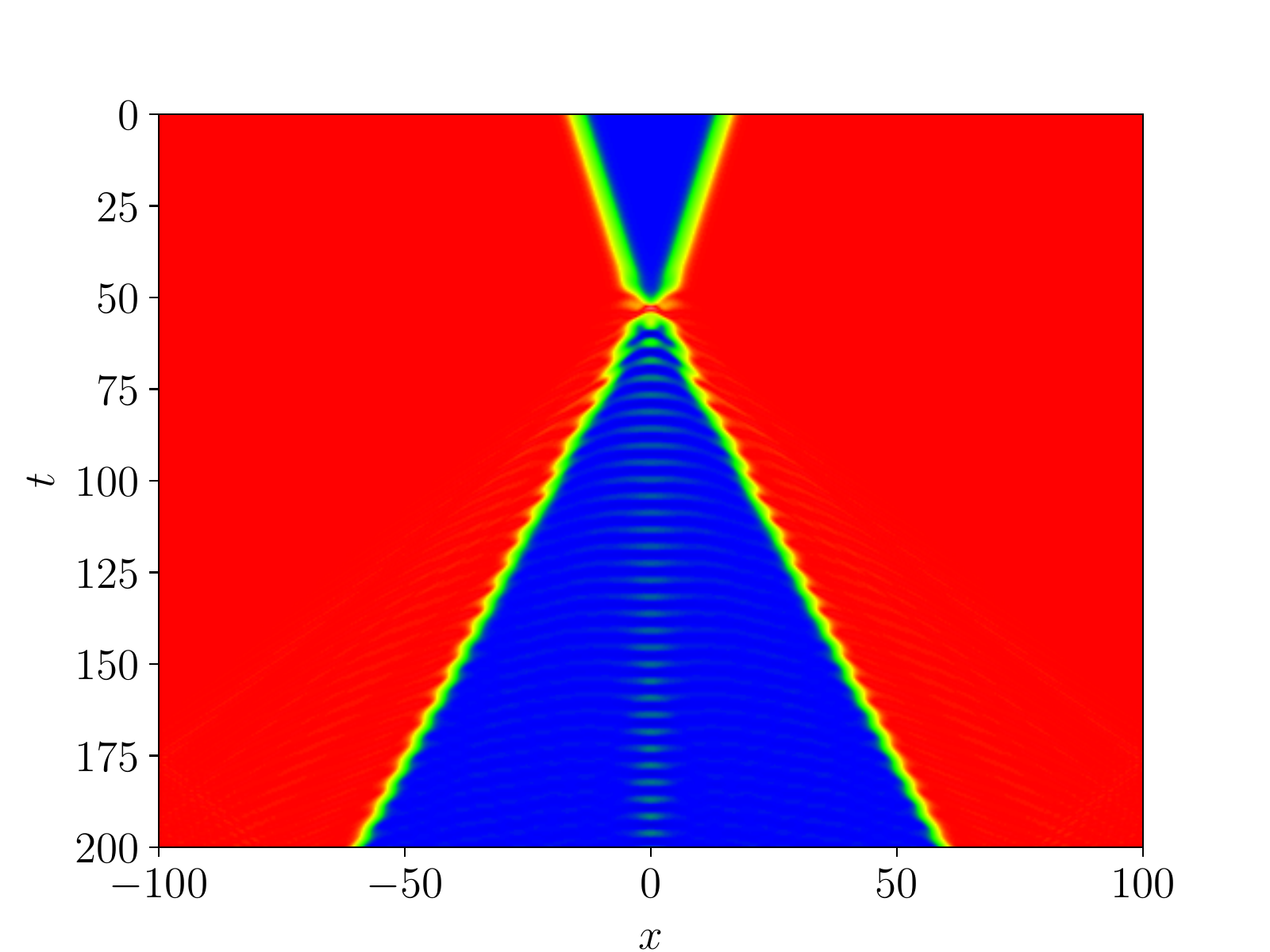}
         \caption{$\phi_0=0.47$, $v_i=0.24$}
     \end{subfigure}
     \begin{subfigure}[b]{0.32\textwidth}         
         \centering
         \includegraphics[width=\textwidth]{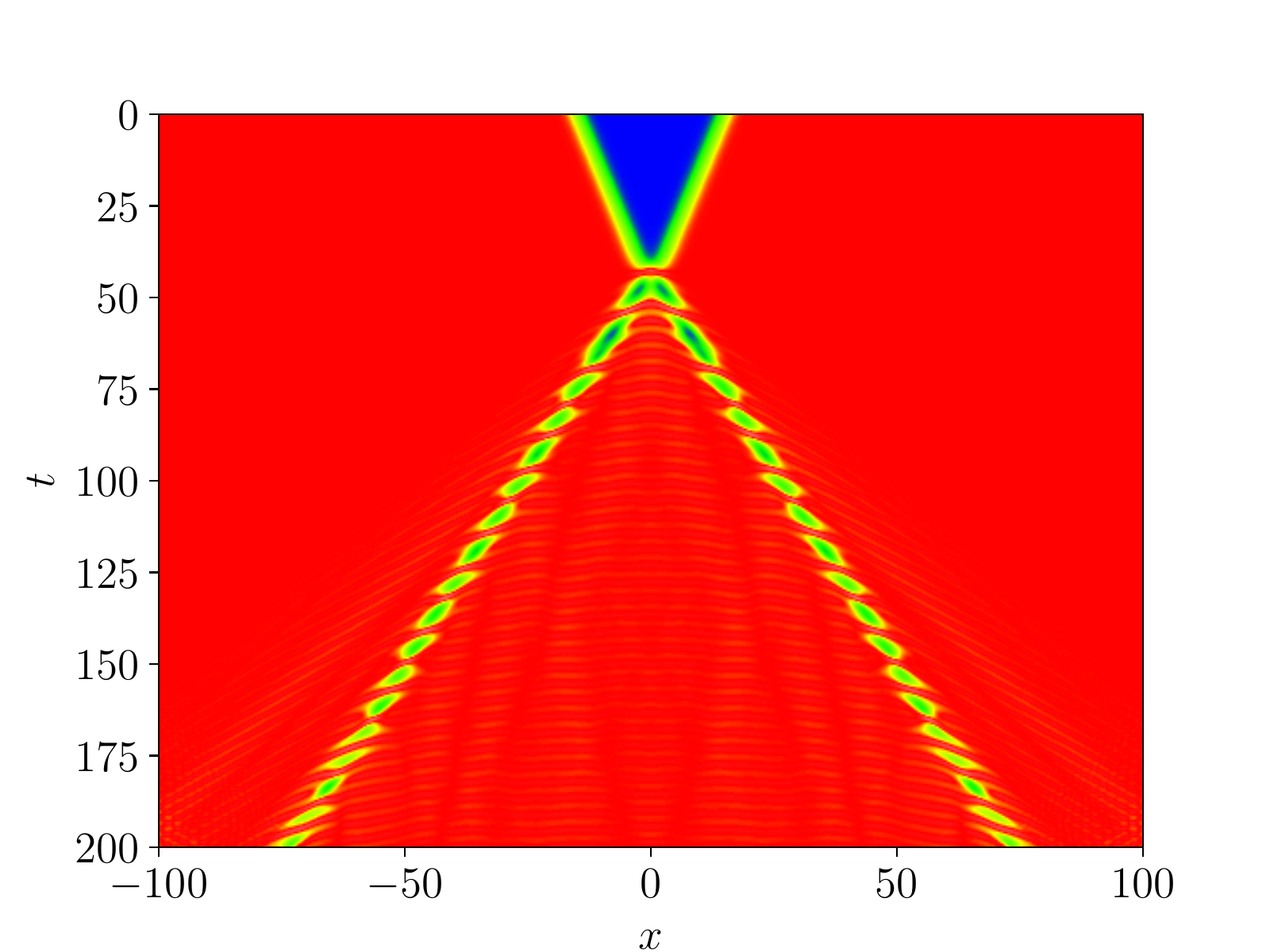}
         \caption{$\phi_0=0.47$, $v_i=0.31$}
     \end{subfigure}
     \begin{subfigure}[b]{0.32\textwidth}         
         \centering
         \includegraphics[width=\textwidth]{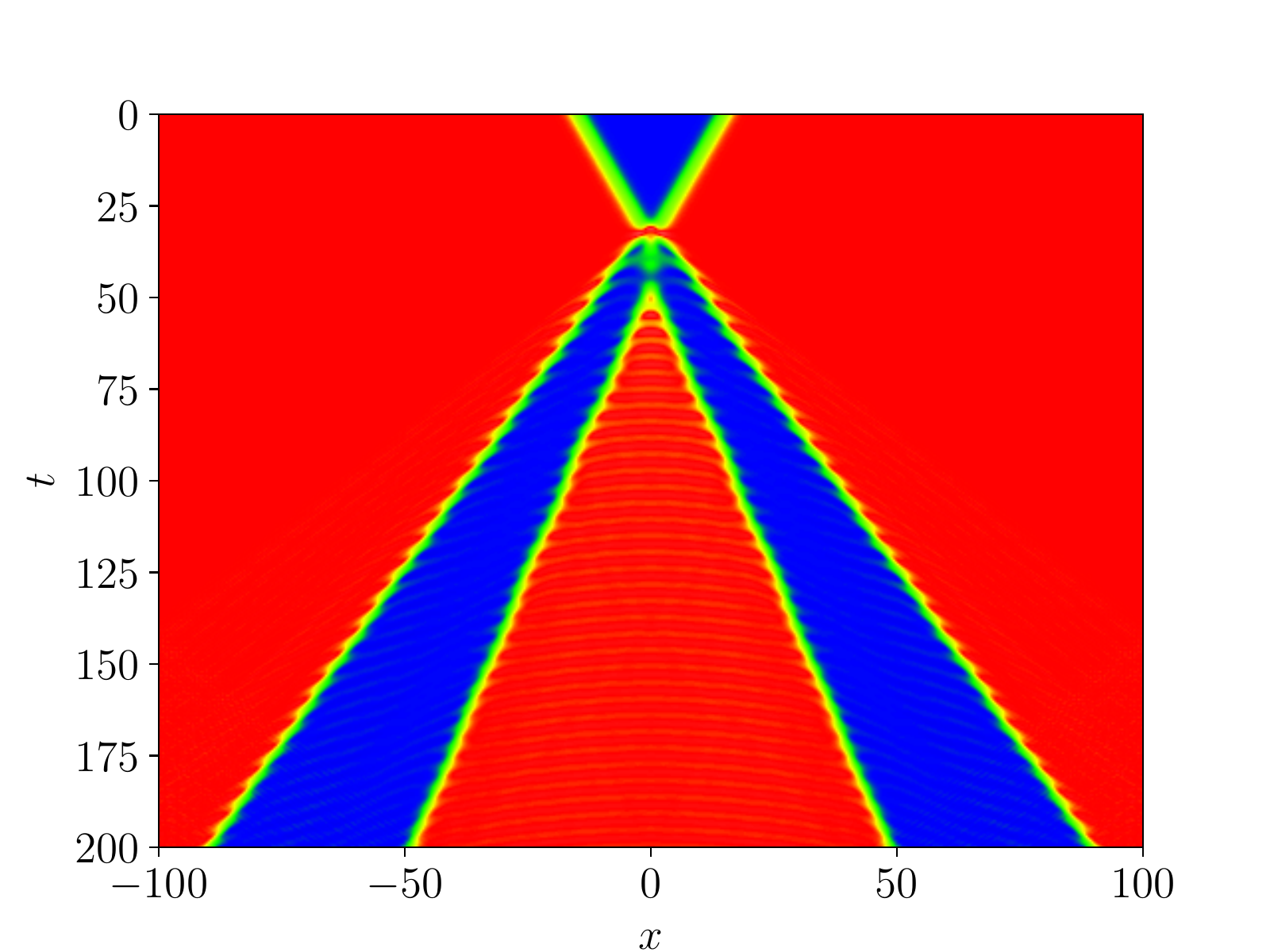}
         \caption{$\phi_0=0.425$, $v_i=0.43$}
     \end{subfigure}
     \caption{Scalar field evolution in spacetime. We consider the case where the fermions start at the excited state.}
   \label{fig_ex}
\end{figure}

For large $\phi_0$, we see a sequence of two-bounce windows at the boundary between the reflection and annihilation regions. As $\phi_0$ is lowered, some resonance windows are suppressed, and others survive. For instance, for $\phi_0=0.76$ and $v_i=0.14$ the resonance window is shown in Fig.~\ref{fig_ex}(a). This corresponds to the third two-bounce window relative to the original $\phi^4$ model. We associate the suppression with the fact that the fermion field does not have a well-defined oscillation between bounces, unlike the one shown in Fig.~\ref{fig_att_cent} for example.

On the far left of Fig.~\ref{fig_mat3}, pseudo-windows and resonance windows alternate. One finds two almost vertical stripes between this region and the yellow and green ones. One stripe is red, and the other is blue, corresponding to two-bounce pseudo- and resonance windows, respectively, with no oscillations between bounces. We illustrate this behavior in Figs.~\ref{fig_ex}(b) and (c). On the left of the stripes, we have an alternating structure of one-bounce pseudo- and resonance windows. They are illustrated in Figs.~\ref{fig_ex}(d) and (e). The boundary between the two behaviors approximately forms a spine structure \cite{campos2021wobbling}, which can be related to the oscillating character of the fermion at the excited state. This also suggests that the fermion is storing energy that can be recovered by the kinks. 

The appearance of one-bounce resonance windows is very similar to what is observed in collisions between wobbling kinks \cite{izquierdo2021scattering}. However, the two results are different. In the present work, the one-bounce resonance windows are separated by pseudo-windows, while for wobbling kinks, they are separated by bions and higher-bounce resonance windows. Even though the model we are dealing with here is supersymmetric and, in principle 
the fermionic excitations could be mapped into the bosonic ones by supersymmetry, the kink-antikink configurations break this symmetry leaving the results quite distinct from wobbling kinks' collisions. 

Finally, at the tip of the red region, we find kink-antikink reflection with the formation of an extra kink-antikink pair. The field evolution, in this case, is shown in Fig.~\ref{fig_ex}(f) as an example. The energy source to create the extra pair comes from the fermion, which becomes more energetic relative to the kink for small $\phi_0$. Let us compute the initial energy contained in the kink-antikink pair to prove this assertion. The energy of a single kink without back-reaction is $E_{cl}=\frac{2\sqrt{2}}{3}\phi_0^2$ \cite{vachaspati2006kinks}. Including a boost and a constant $C$, which is different from unity due to the fermion back-reaction, we obtain 
\begin{equation}
E_2=C\frac{4\sqrt{2}}{3\sqrt{1-v_i^2}}\phi_0^2.
\end{equation} 
for the kink-antikink pair with $C\approx1.166$ which we find numerically. Plugging in $v_i=0.43$ and $\phi_0=0.425$, we obtain $E_2\approx0.440$, which is much less than the minimum energy to create two kink-antikink pairs at rest $E_{4,min}=\frac{8\sqrt{2}}{3}\phi_0^2\approx0.681$.

The behavior of fermions during collisions is very similar to what we found in the previous cases. After the collision, a significant fraction of the fermion density stays bound to the kinks, the oscillons, or the bion. Moreover, oscillations with a nonlinear character are also observed for small $\phi_0$.

\section{Conclusion}

In this work, we studied a supersymmetric model containing fermion-kink interactions. We showed that the equations of motion contain fermion back-reaction to the kink. The back-reaction becomes more relevant for light kinks, which possess a small value at infinity, $\phi_0$. 

We computed both zero-energy and excited fermion bound states and used boosted configurations to construct symmetric collisions. We have shown that two parameters could control the strength of the fermion back-reaction on the kinks in a symmetric collision, the parameter $\phi_0$ and the phase between the fermions on the kink and antikink $A$. We focused on cases with maximum back-reaction, $A=\pm 1$ for a fixed value of $\phi_0$. Moreover, we have estimated the force between the fermions computing it from the energy-momentum tensor. The force can be either attractive or repulsive. As $\phi_0$ is lowered, the interaction between the fermions can become larger than the scalar field attraction.

Then, we studied the stability equation of the kink-fermion system. The stability equation is given by a complicated linear operator, which needs to be analyzed carefully. We employed a simple discretization method and then diagonalized the matrix numerically. For the composite system with the fermion at the zero mode, we found a spectrum containing two modes with eigenvalue $\lambda=0$, which can be found analytically, and degenerate excited states. Moreover, the analysis suggests that the composite system is stable in all cases because we did not find positive and real eigenvalues in our numerical analysis.

Next, we analyzed the equations of motion and discussed the necessary condition to have a non-vanishing back-reaction. We integrated the equations of motion in time and found a wide variety of behaviors expected by the back-reaction's nonlinear character. For the attractive case with the fermion at the zero mode, we observed bion formation, resonance windows, reflection, and pseudo-windows, where two oscillons are formed. We also observed reflection without contact or with a soft collision for the repulsive case. In both cases, we found a rich structure of resonance windows up to very low values of $\phi_0$. 

After colliding, a significant fraction of the fermion density is bound at the localized structures in the scalar field. Moreover, the density shows complicated oscillation patterns that a linear combination of vibrational modes cannot describe. Interestingly, we observed that the fermions exhibit the same number of oscillations between bounces as the scalar field. However, this only occurs for the zero-mode initial condition.

Many fascinating behaviors appeared in the collision between composite systems with excited fermions. First, some resonance windows are suppressed as $\phi_0$ is lowered. This seems to be related to fact that the resonant behavior of the fermion field shown in Fig.~\ref{fig_att_cent} does not occur anymore. For even lower values of $\phi_0$, we found an alternating structure of pseudo-windows and one-bounce resonance windows related to the oscillating character of the excited fermion states. This is similar to what happens in collisions between wobbling kinks \cite{izquierdo2021scattering, campos2021wobbling}, but in this case, the oscillation is performed by the fermion. Therefore, it appears that the fermion is storing part of the resonant energy, but a more detailed analysis is needed, as will be discussed below.

The resonant energy exchange mechanism is clearly in play because resonance windows are observed. For large $\phi_0$, the exchange is conducted by the kink's shape mode. The reason is that the back-reaction is negligible, and we recover the original $\phi^4$ in this limit. When $\phi_0$ is lowered, it is natural to ask if the fermion excitation also conducts the energy exchange. Even though we found evidence that this is the case, we cannot give a definite answer in the present work because both fermion and kinks excitations have the same frequency, and it is not possible to distinguish between them. It would also be interesting to investigate whether a fermion excited state alone could create a resonant structure. This and other pending issues discussed in this section are the subjects of an ongoing investigation.

\section*{Acknowledgments}

We acknowledge financial support from the Brazilian agencies CAPES and the National Council for Scientific and Technological Development - CNPq. AM also thanks financial support from Universidade Federal de Pernambuco Edital Qualis A. We thank Mauro Copelli for the access to his lab's computer cluster, which was essential for conducting the research reported in this paper. Part of the simulations was also conducted in the SDumont cluster from the Brazilian laboratory LNCC (Laborat\'orio Nacional de Computa\c{c}\~ao Cient\'ifica).

\end{document}